\documentclass[prb,manuscript,aps,showpacs]{revtex4-1}
\usepackage{graphicx}
\newcommand{\om}{\omega}

\newcommand{\E}{{\bf E}}
\newcommand{\ev}{{\bf e}}

\newcommand{\K}{{\bf K}}
\newcommand{\rv}{{\bf r}}
\newcommand{\R}{{\bf R}}

\newcommand{\Hv}{{\bf H}}

\newcommand{\pv}{{\bf p}}

\newcommand{\uv}{{\bf u}}
\newcommand{\xu}{\hat{{\bf x}}}
\newcommand{\yu}{\hat{{\bf y}}}
\newcommand{\zu}{\hat{{\bf z}}}

\newcommand{\shat}{\hat{s}}
\newcommand{\ppp}{\hat{p}^+}
\newcommand{\ppm}{\hat{p}^-}

\newcommand{\green}{{\stackrel{\leftrightarrow}{\bf G}}}
\newcommand{\iden}{{\stackrel{\leftrightarrow}{\bf I}}}
\newcommand{\hperp}{{\stackrel{\leftrightarrow}{\bf h}}}
\newcommand{\gperp}{{\stackrel{\leftrightarrow}{\bf g}}}
\newcommand{\gamperp}{{\stackrel{\leftrightarrow}{\bf \Gamma}}}
\newcommand{\deter}{{\stackrel{\leftrightarrow}{\bf D}}}
\newcommand{\dyadic}{{\stackrel{\leftrightarrow}{\bf A}}}
\newcommand{\dyadicb}{{\stackrel{\leftrightarrow}{\bf B}}}
\newcommand{\dyadicc}{{\stackrel{\leftrightarrow}{\bf C}}}

\newcommand{\alpht}{{\stackrel{\leftrightarrow}{\bf \alpha}}}
\newcommand{\mv}{{\bf m}}

\begin{document}

\title{Tip-sample electromagnetic interaction in the infrared: Effective polarizabilities, retarded image dipole model and near-field thermal radiation detection}
\author{Karl Joulain}
\affiliation{Institut PPrime, CNRS-Universit\'e de Poitiers-ENSMA, UPR 3346, 2, Rue Pierre Brousse, B.P 633, 86022 Poitiers Cedex, France.}
\author{P. Ben-Abdallah}
\affiliation{Laboratoire Charles Fabry, UMR 8501, Institut d'Optique, CNRS, Universit\'e Paris-Sud 11, 2 avenue Augustin Fresnel, 91127 Palaiseau Cedex, France}
\author{P.-O. Chapuis}
\affiliation{Catalan Institute of Nanotechnology (ICN), 
Campus UAB, 08193 Bellaterra (Barcelona), Spain}
\affiliation{Centre de Thermique de Lyon (CETHIL), 
CNRS, INSA de Lyon, UCBL, Campus de la Doua, 69621 Villeurbanne, France}
\author{Y. De Wilde}
\affiliation{Institut Langevin, ESPCI Paris Tech, 10 rue Vauquelin, 75005 Paris, France}
\author{A. Babuty}
\affiliation{Institut Langevin, ESPCI Paris Tech, 10 rue Vauquelin, 75005 Paris, France}
\date{Monday, February 6$^{th}$ 2012}
\begin{abstract}
We analyse how a probing particle modifies infrared electromagnetic near fields. The particle, assimilated to both electric and magnetic dipoles, represents the tip of an apertureless scanning optical near-field microscope (SNOM). We show that the interaction can be accounted for by ascribing to the particle effective dipole polarizabilities that add the effect of retardation to the one of the image dipole. Apart from these polarizabilities, the SNOM signal expression depends only on the fields without tip perturbation, shown to be closely related to the electromagnetic density of states (EM-LDOS) and essentially linked to the sample's optical properties, so that measuring local spectra of heated samples is equivalent to performing a local surface spectroscopy. We also analyse the case where the probing particle is hotter. We evaluate in this case the impact of the effective polarizabilities on the tip-sample near-field radiative heat transfer. We also show that such an heated probe above a surface also performs a surface spectroscopy. The calculations agree well with available experimental data. 
\end{abstract}

\maketitle

\section{Introduction}
Since the seminal works of Rytov \cite{Rytov:1989ur}, it is known that thermal radiation has a different behaviour when the involved caracteristic lengths are large or small compared to the thermal wavelength \cite{Joulain:2005ih,Volokitin:2007el,Dorofeyev:2011bg}. For example,
the heat flux transferred between bodies separated by
a subwavelength distance can exceed by far the one between perfect blackbodies \cite{Polder:1971uu,BenAbdallah:2010hp}. Energy density \cite{Shchegrov:2000td} and coherence properties \cite{Henkel:2000tr} are also strongly affected in the near-field, especially close to materials exhibiting resonances such as polaritons. Knowing precisely how the electromagnetic field behaves close to a surface is therefore an important issue in order to address all potential applications involving near-field heat transfer.

From an experimental point of view, near-field radiation  coherence properties have been utilized to produce directional and monochromatic thermal sources \cite{Greffet:2002ur,Lee:2006cj,Biener:2008cj}. More recently, near-field radiative heat transfer enhancement have been detected by means of probe microscopy techniques \cite{Kittel:2005fr,Narayanaswamy:2008gj,Rousseau:2009es,Ottens:2011kh}. Near-field thermal flux imaging has been operated with a scanning thermal microscope \cite{Kittel:2008bc,Wischnath:2008hp}. A scanning near-field optical microscopy (SNOM) without external enlightment, termed thermal radiation scanning tunneling microscopy (TRSTM), has also been shown to be useful to image surfaces \cite{DeWilde:2006kt} and to be very sensitive \cite{Kajihara:2010fo,Kajihara:2011uu}. Very recently, local spectra has also been performed \cite{Babuty:dYmzb9en}.
All these experimental techniques use a small probe brought to the vicinity of the studied material surface. When the probe is approached close to the surface, mutual interaction between the tip and the surface modifies the local electromagnetic field\cite{GarciadeAbajo:2007eb,Intravaia:2010gp,BenAbdallah:2011be,Castanie:2011ue}, a phenomenon described by probe optical properties such as its polarizability if it is modelled as a dipole. 
This is the origin of the field scattering to the far field, which is beneficial for TRSTM, but also complicates the data analysis for the previoulsy-mentioned near-field techniques although one can relate the local electromagnetic field to the surface material optical properties. 
A fundamental issue is to know how the 
TRSTM signal scattered by a tip and detected in the far-field can be related to the electromagnetic local density of states (LDOS)  \cite{Joulain:2003hc,Kittel:2008bc}.  Is there a way that a SNOM detecting thermal radiation might be the electromagnetic equivalent of the scanning tunneling microscope detecting electronic LDOS \cite{Tersoff:1985wm}? Moreover, when a probe is heated, one can ask itself how fast it cools down and what the signal detected in far-field is when it is approached close to a surface.

If some of these questions have already been addressed in the past \cite{Joulain:2003hc,Mulet:2001kp}, our goal is here to clarify remaining interrogations and to complete the body of work.  Following previous similar works \cite{Knoll:2000wm,Sun:2007cl}, we will first see how the particle polarizability 
can be replaced by an effective polarizability taking into account the multiple reflections between the probe and the surface. %
This model is more general than the image-dipole one \cite{Knoll:2000wm} whose range of validity is very restricted in the infrared.
 We will then use the theory to calculate the signal detected in the far-field due to scattering of the near field by a probe. 
We will simulate 
the SNOM signal obtained 
by scanning 
a surface excited either by a plasmon or more generally by 
thermal excitation (near-field thermal emission). 
We will consider in the following two parts 
that the probe tip can be assimilated to both electric and magnetic dipoles. We will first 
give the signal detected in far field 
due to a heated 
probe 
when accounting for the surface reflections.
We will show then how 
cooling is increased 
when the particle is approached close to the surface.

\section{Expressions of the effective polarizabilities}
\begin{figure}[h]
\centering
\includegraphics[width=10cm]{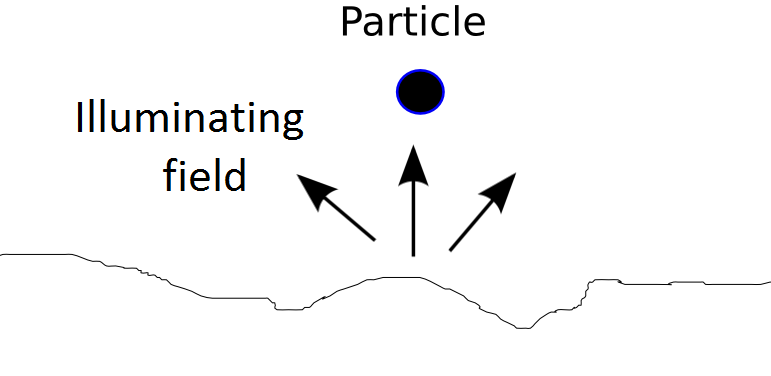}
\caption{System scheme}
\label{systeme}
\end{figure}

We propose here to calculate the effective polarizability of a dipolar particle when it is placed in an environment which is different from vacuum. Indeed, when a particle is added in a system, the electromagnetic field present in the system illuminates the particle, what gives birth to an induced dipole moment centered at its position. This dipole radiates also a field everywhere, including at the particle position. Interactions between the particle and the system modify the total electromagnetic field, which becomes then different from the one in absence of a perturbating particle. Our aim is to show that we can consider the electromagnetic field to be unchanged at the particle's center if we ascribe to the particle an effective polarizability that accounts for the perturbation (Fig. \ref{systeme}). 

Let us call $\E^0$ the electromagnetic field in the system in the absence of particle. When a particle is placed in the system, an electric dipole $\pv$ and a magnetic dipole $\mv$ will appear at the position of the particle $\rv_p$. These dipoles radiate a field : the total local field $\E^{loc}$ is the sum of the field in the absence of particle with the field radiated by the dipoles. In the following, we write the dipole radiated field with the Green's tensor help.
\begin{equation}
\label{ }
\E^{loc}(\rv)=\E^{0}(\rv)+\green^{EE}(\rv,\rv_p)\cdot\pv+\green^{EH}(\rv,\rv_p)\cdot\mv
\end{equation}
that can also be written
\begin{equation}
\label{eloc}
\E^{loc}(\rv)=\E^{0}(\rv)+\green^{EE}(\rv,\rv_p)\cdot\alpha^E\E^{loc}(\rv_p)+\green^{EH}(\rv,\rv_p)\cdot\alpha^H\Hv^{loc}(\rv_p)
\end{equation}
where $\alpha^E$ and $\alpha^H$ are the particle electric and magnetic dipoles when it is alone in vacuum.

Analogous expressions exist for the magnetic field
\begin{equation}
\label{}
\Hv^{loc}(\rv)=\Hv^{0}(\rv)+\green^{HE}(\rv,\rv_p)\cdot\pv+\green^{HH}(\rv,\rv_t)\cdot\mv
\end{equation}
that can also be written
\begin{equation}
\label{hloc}
\Hv^{loc}(\rv)=\Hv^{0}(\rv)+\green^{HE}(\rv,\rv_p)\cdot\alpha^E\E_{loc}(\rv_p)+\green^{HH}(\rv,\rv_p)\cdot\alpha^H\Hv^{loc}(\rv_p)
\end{equation}
The Green tensors used here are of four type : $\green^{EE}(\rv,\rv_p)$ is the Green tensor that gives the electric field at position $\rv$ when an electric dipole is placed at $\rv_p$. In the same way, $\green^{EH}(\rv,\rv_p)$ is the Green tensor that gives the electric field at position $\rv$ when an magnetic dipole is placed at $\rv_p$.  $\green^{HE}$ and $\green^{HH}$ respectively gives the magnetic field of an electric and a magnetic dipole. These Green tensors, which can all be expressed in terms of $\green^{EE}(\rv,\rv_p)$ \cite{Joulain:2010bq}, are the sum of a direct contribution (that is the Green tensor in vacuum) and of a perturbed contribution.
When one considers the electromagnetic field at the particle position $\rv_p$, only the perturbed Green tensor contributes. Then, the local field is obtained solving the system (\ref{eloc}) (\ref{hloc}). 

\begin{eqnarray}
\E^{loc}(\rv_p) & = &\dyadic^{-1} \E^0(\rv_p)+\dyadicb^{-1}\Hv^0(\rv_p) \label{elocfull} \\
\Hv^{loc}(\rv_t) & = & \dyadicc^{-1} \E^0(\rv_p)+\deter^{-1}\Hv^0(\rv_p)\label{hlocfull} 
\end{eqnarray}
where 
\begin{eqnarray}
\dyadic & = & [\iden-\alpha^E\green^{EE}(\rv_p,\rv_p)]-\alpha^H\green^{EH}(\rv_p,\rv_p)[\iden-\alpha^H\green^{HH}(\rv_p,\rv_p)]^{-1}\alpha^E\green^{HE}(\rv_p,\rv_p) \\
\dyadicb & = & - \alpha^E\green^{HE}(\rv_p,\rv_p)+[\iden-\alpha^H\green^{HH}(\rv_p,\rv_p)][\alpha^H\green^{EH}(\rv_p,\rv_p)]^{-1}[\iden-\alpha^E\green^{EE}(\rv_p,\rv_p)]\\
\dyadicc & = & - \alpha^H\green^{EH}(\rv_p,\rv_p)+[\iden-\alpha^E\green^{EE}(\rv_p,\rv_p)][\alpha^E\green^{HE}(\rv_p,\rv_p)]^{-1}[\iden-\alpha^H\green^{HH}(\rv_p,\rv_p)]\\
 \deter& = & [\iden-\alpha^H\green^{HH}(\rv_p,\rv_p)]-\alpha^E\green^{HE}(\rv_p,\rv_p)[\iden-\alpha^E\green^{EE}(\rv_p,\rv_p)]^{-1}\alpha^H\green^{EH}(\rv_p,\rv_p)
\end{eqnarray}
The preceding equations are general for the calculation of a local field at dipole position. 
Apart from the fact that magnetic dipole is taken into account, the self-consistent reasoning used to obtain local fields is very similar to previous work using the so-called coupled dipole theory such as \cite{Lax:1951tb,GarciadeAbajo:2007eb,Intravaia:2010gp,BenAbdallah:2011be}. Local fields only depend on the system Green's tensors. Particle dipoles can be related to the field at the tip position in the absence of particle
\begin{equation}
\label{ }
\left(\begin{array}{c}
      \pv   \\
      \mv
\end{array}\right)=\left(\begin{array}{cc}
    \alpha^E\dyadic  & \alpha^H\dyadicb    \\
    \alpha^E\dyadicc  &   \alpha^H\deter
\end{array}\right)\left(\begin{array}{c}
      \E^0  \\
      \Hv^0
\end{array}\right)
\end{equation}
so that the four elements in the matrix can be seen as effective polarizability expressions.

We now highlight the case of a simple system made of a planar interface separating a material and vacuum. In this situation, Green tensors are well known \cite{Sipe:1987td,Joulain:2010bq} and can be written knowing the materials optical properties. We recall the direct Green's tensor describing propagation in vacuum
\begin{eqnarray}
\green^{AB}_0(\rv,\rv',\om)&=&\frac{ik_0^2C_{AB}}{8\pi^2}\int\frac{d^2\K}{\gamma}(\hat{e_A^{s-}}\hat{e_B^{s-}}+\hat{e_A^{p-}}\hat{e_B^{p-}})e^{i\K.(\R-\R')}e^{-i\gamma(z-z')} 
\end{eqnarray}
and the one associated to a plane wave reflection at an interface
\begin{eqnarray}
\green^{AB}_R(\rv,\rv',\om)&=&\frac{ik_0^2C_{AB}}{8\pi^2}\int\frac{d^2\K}{\gamma}(\hat{e_A^{s+}}r^s\hat{e_B^{s-}}+\hat{e_A^{p+}}r^p\hat{e_B^{p-}})e^{i\K.(\R-\R')}e^{i\gamma(z+z')}
\end{eqnarray}
Here $K$ and $\gamma$ are respectively the parallel and the perpendicular component of the wavevector. $r^s$ and $r^p$ are the Fresnel reflection coefficients for polarizations $s$ and $p$ which depend only on the material optical properties and $K$.
Introducing vectors $\shat$ and $\ppm$ vectors used by Sipe \cite{Sipe:1987td} :
\begin{eqnarray}
\shat & = & \K\times\zu/K \\
\ppm & = & \frac{1}{k_0K}\left[K^2\zu+\gamma K_x\xu+\gamma K_y\yu\right] \\
\ppp & = & \frac{1}{k_0K}\left[K^2\zu-\gamma K_x\xu-\gamma K_y\yu\right] \\
\end{eqnarray} 
one writes the vectors in Green dyadics as
\begin{eqnarray}
\label{ }
\hat{e_E^{s+}}=\hat{e_E^{s-}}=\hat{e_H^{p+}}=\hat{e_H^{p-}}=\shat\\
\hat{e_E^{p+}}=-\hat{e_H^{s+}}=-\ppp\\
\hat{e_E^{p-}}=-\hat{e_H^{s-}}=-\ppm
\end{eqnarray}
Moreover, $C_{EE}=\epsilon_0^{-1}$, $C_{EH}=\mu_0c$, $C_{HE}=\epsilon_0c$, $C_{HH}=1$.
When one considers the electromagnetic field at the particle position $\rv_p$, only the reflected Green's dyadic contributes. In this plane-parallel geometry, $\green^{EH}_R(\rv_p,\rv_p)=\green^{HE}_R(\rv_p,\rv_p)=0$. 
Expressions of the electromagnetic field at the tip position can be obtained from simplification of (\ref{elocfull}) and (\ref{hlocfull}) taken in $\rv=\rv_p$. Then
\begin{eqnarray}
\E^{loc}(\rv_p) & = &\left[\iden-\alpha^E\green^{EE}_R(\rv_p,\rv_p)\right]^{-1}\E^0(\rv_p)  \\
\Hv^{loc}(\rv_p) & = & \left[\iden-\alpha^H\green^{HH}_R(\rv_p,\rv_p)\right]^{-1}\Hv^0(\rv_p) 
\end{eqnarray}
leading to the following dipole expressions :
\begin{eqnarray}
\pv(\rv_p) & = & \alpha^E\left[\iden-\alpha^E\green^{EE}_R(\rv_p,\rv_p)\right]^{-1}\E^0(\rv_p)  \\
\mv(\rv_p)& = & \alpha^H \left[\iden-\alpha^H\green^{HH}_R(\rv_p,\rv_p)\right]^{-1}\Hv^0(\rv_p) 
\end{eqnarray}
We note that we can recast these formulae in order to directly relate the dipole moments and the fields in absence of dipole by introducing the following effective polarisabilities
\begin{equation}
\label{alpheeff}
\alpht^E=\alpha^E\left[\iden-\alpha^E\green_{EE}^R(\rv_t,\rv_t)\right]^{-1}
\end{equation} 
and 
\begin{equation}\
\alpht^H=\alpha^H\left[\iden-\alpha^H\green_{HH}^R(\rv_t,\rv_t)\right]^{-1}
\end{equation}
These effective polarizabilities are anisotropic and account for the fields multiple reflections between the particle and the interface.

\section{Parametric study of the effective polarizabilities}

In this section, we analyze how the effective polarizabilties vary as a function of the material optical properties, particle sizes and particle-surface distance. We consider that the particles can be assimilated to spherical dipoles, with well-established "vacuum" polarizabilities. Note that these ones depend, independently from the influence of the interface studied here, 
on the ratio of the particle radius $R_p$ to the wavelength $\lambda$ (Mie parameter$ x=2\pi R_p / \lambda$) and also on  $y=\sqrt{\epsilon}x$, where $\epsilon$ is the particle dielectric constant. The following expressions  given by Chapuis \cite{Chapuis:2008kca} apply if the wavelength is much larger than $R_p$: 
\begin{equation}
\label{ }
\alpha_E(\omega)=\epsilon_02\pi R_p^3\frac{2\left[\sin(y)-y\cos(y)\right]-x^2\left[\frac{-\sin(y)}{y^2}+\frac{\cos(y)}{y}+\sin(y)\right]}{\left[\sin(y)-y\cos(y)\right]+x^2\left[\frac{-\sin(y)}{y^2}+\frac{\cos(y)}{y}+\sin(y)\right]}
\end{equation}
and 
\begin{equation}
\label{ }
\alpha_H(\omega)=-2\pi R_p^3\left[\left(1-\frac{x^2}{10}\right)+\left(-\frac{3}{y^2}+\frac{3}{y}\cot (y)\right)\left(1-\frac{x^2}{6}\right)\right]
\end{equation}

Our aim is to compare the theoretical results to apertureless Scanning Near-field Microscopy (SNOM) results. We consider particles made of tugnsten, as this material is often used for SNOM tips. We study the effective polarisabilities for two spherical tip sizes (100nm and 1 $\mu$m radii) and above three materials : SiC and SiO2, being both dielectrics, and gold, a metal. 

\subsection{Different contributions to the effective polarizabilities}

Let us consider the case of a 100nm-radius tip of tungsten situated 100 nm above an interface separating the vacuum and the sample. This corresponds to a sphere in contact with the interface. In principle, the dipole model is not sufficient anymore to describe this situation since the evanescent field is not constant inside the particle and therefore multipolar contributions should be considered. However, due to the complexity of this task, the use of dipoles to model the tip has been the rule for several SNOM theories and will be considered as a limiting case here. It shows most of the modifications that occur when the tip is approached close to a surface.
 
\begin{figure}
\begin{center}
\includegraphics[width=12cm]{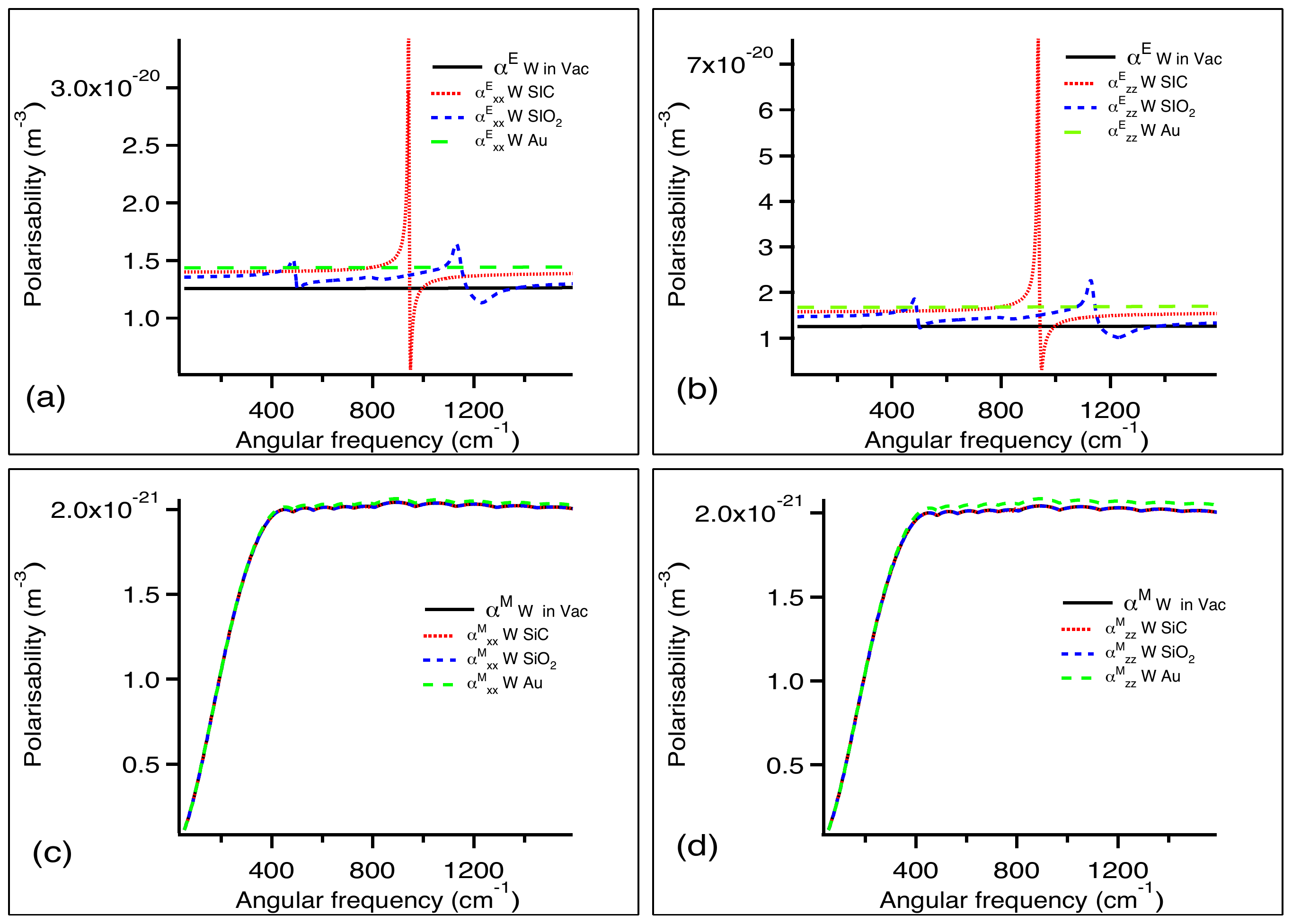}
\caption{Parallel electric (a), perpendicular electric (b), parallel magnetic (c) and perpendicular magnetic (d) polarizability for a 100nm-radius spherical tungsten tip situated at $z$=100 nm above an plane suface of SiC, SiO2 or Gold.}
\label{aleff100100}
\end{center}
\end{figure}

In the case of a dipolar particle close to a plane interface, four different contributions to the polarizabilities are identified: The parallel electric ($\alpha^E_{xx}$), perpendicular electric ($\alpha^E_{zz}$), parallel magnetic ($\alpha^H_{xx}$) and perpendicular magnetic ($\alpha^H_{zz}$) ones. They are plotted in Fig. \ref{aleff100100}.
 
We observe that both perpendicular and parallel effective electrical polarisabilities are very different from the vacuum polarisability of the single particle. In particular, resonances are observed respectively close to the SiC and SiO$_2$ plasmon polariton resonance frequencies. This is not surprising since the single-interface Green tensors enter in the expresions of the effective polarizabilities. Since the dyadics depend strongly on the reflection coefficients, the former take large values close to the resonance frequencies, especially at small distances from the interface. Thus the effective polarizabilities are greatly affected at small distances and close to resonant frequencies. In contrast, no peak is seen in the spectra of gold effective polarizabilities since gold does not exhibit resonances in the studied frequency range. Thus, the effective polarizabilities are different from the vacuum ones as the presence of evanescent modes close to the interface affect the particle.
The observation of the effective magnetic polarizaties spectra show very different variations from the electric one. The main reason is that vacuum magnetic polarizabilities have different asymptotic variations at small radii $R_p$. Indeed, the electrical polarizabilities behave in the small-particle regime as $R_p^3$  (Clausius-Mossotti limit) whereas the magnetic ones behave as $R_{p}^{5}$ $ k_{0}^{2} \left(\epsilon -1 \right)$ \cite{Chapuis:2008kcb}. In the present case, as the particle radius is much smaller than the wavelength, the magnetic polarizability is smaller, which implies that the correction is also smaller. We note that the correction is larger for gold than for SiC and SiO$_2$. This is a consequence of the fact that evanescent magnetic modes are more present above metals than above dielectrics, as shown by the large value of the metal dielectric constant in the infrared.
  
\begin{figure}
\begin{center}
\includegraphics[width=12cm]{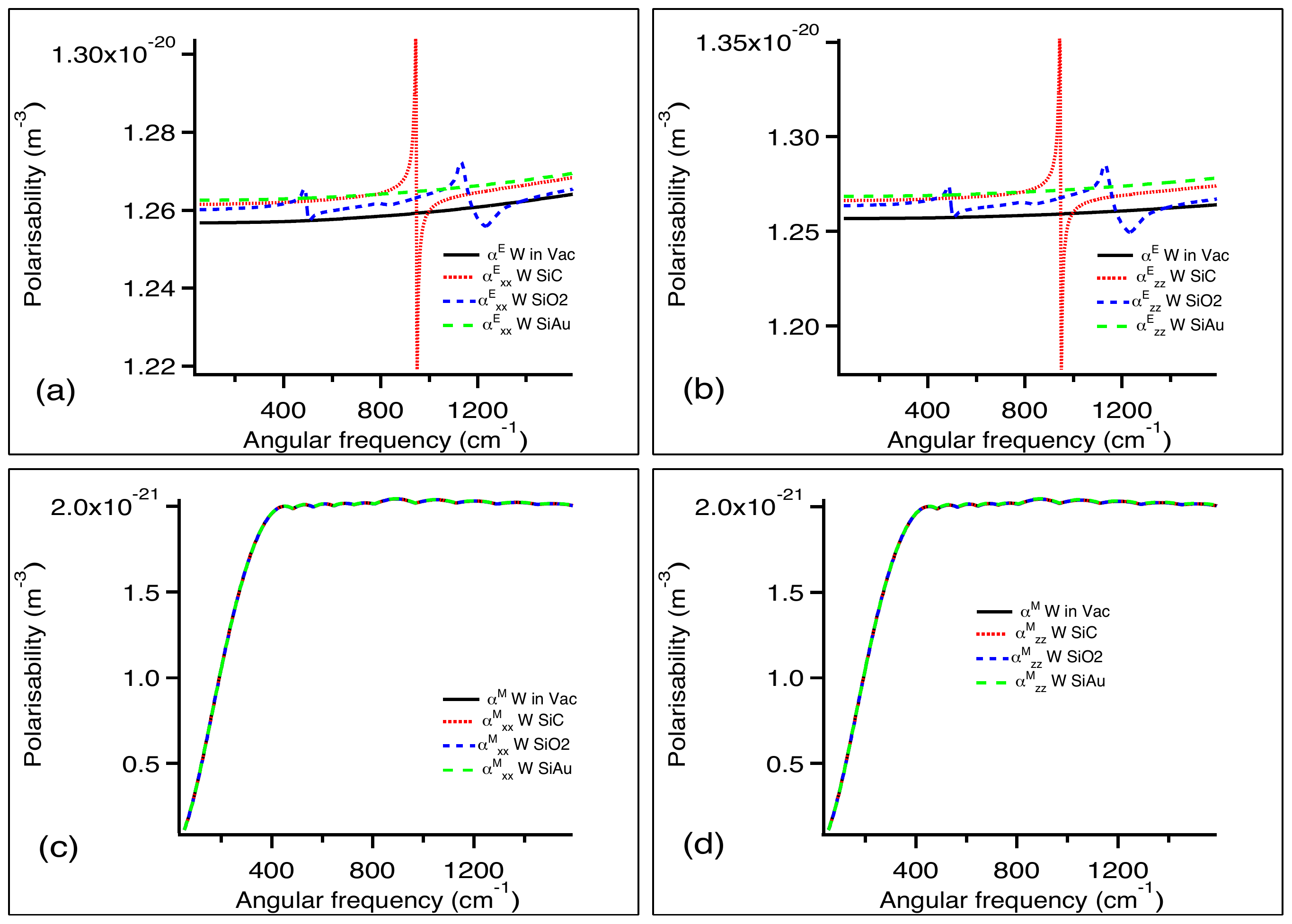}
\caption{Parallel electric (a), perpendicular electric (b), parallel magnetic (c) and perpendicular magnetic (d) polarizability for a 100nm-radius spherical tungsten tip situated at $z$=300 nm above an plane suface of SiC, SiO2 or Gold.}
\label{aleff100300}
\end{center}
\end{figure}

The same particles polarizabilities have been studied at a larger distance to the interface i.e for 300 nm. The variations are shown in Fig. \ref{aleff100300}. The electrical polarizabilities spectrum behaviour with variations around resonant frequencies is still present at larger distance although the corrections to the vacuum polarizability are much smaller. Thus, evanescent modes decay exponentially when the distance to the interface increases. This fact also explains why corrections are almost inexistant at $z$=300 nm for magnetic polarizability. 

\begin{figure}
\begin{center}
\includegraphics[width=12cm]{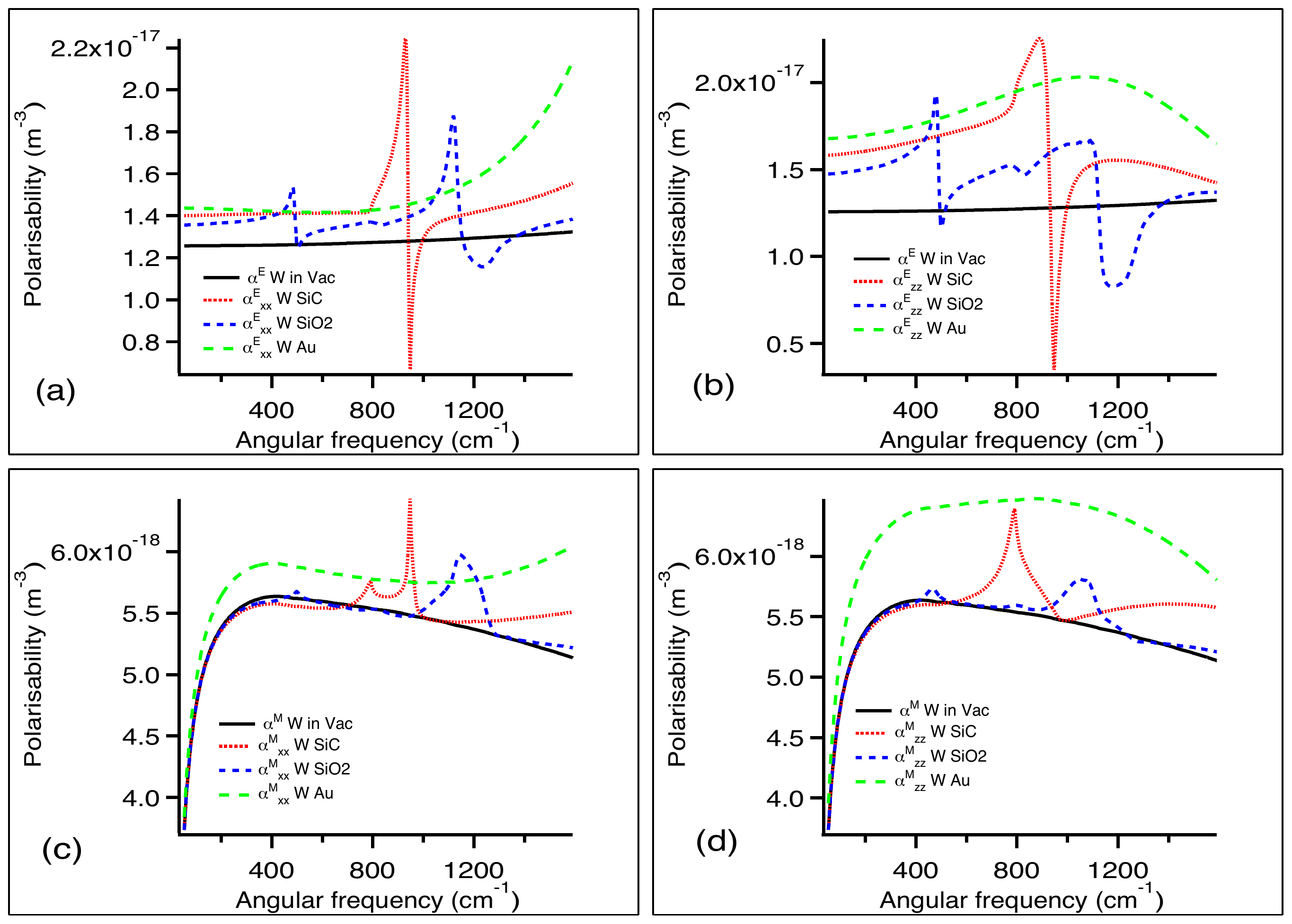}
\caption{Parallel electric (a), perpendicular electric (b), parallel magnetic (c) and perpendicular magnetic (d) polarizability for a 1$\mu$m-radius spherical tungsten tip situated at $z$=1 $\mu$m above an plane suface of SiC, SiO2 or Gold.}
\label{aleff1mic1mic}
\end{center}
\end{figure}

Let us now study the case of a larger particle (1 $\mu$m) situated at 1 $\mu$m form the interface. The effective polarizabilities are represented in Fig. \ref{aleff1mic1mic}.
We note important corrections for both effective electric and magnetic polarizabilities. These corrections are once again much more important around surface resonance frequencies.
The impact on the magnetic polarizabilities is now much more striking than for small particles due to the fact that here $k_0 R_p$ is larger. We will see in the next section that retardation effects begin to enter into play at such distances so that corrections at 1$\mu$m are different from the one at 100 nm as can be noticed when one analyzing carefully the effective polarizabilities correction curves.

\subsection{Effective polarizability asymptotic expression}
We now study the effetive polarizabilities in order to obtain asymptotic expressions in some limited physical situations. We first start with Green tensors that appear as an integral over the parallel wavevector:

\begin{equation}
\label{ }
\green^{EE}_R(\rv_t,\rv_t)=\frac{i\mu_0\omega^2}{4\pi}\left(\begin{array}{ccc}\int_0^\infty\frac{KdK}{2\gamma}(r^s-r^p\frac{\gamma^2}{k_0^2})e^{2i\gamma z_t} & 0 & 0 \\0 & \int_0^\infty\frac{KdK}{2\gamma}(r^s-r^p\frac{\gamma^2}{k_0^2})e^{2i\gamma z_t} & 0 \\0 & 0 & \int_0^\infty\frac{K^3dK}{\gamma k_0^2}r^pe^{2i\gamma z_t}\end{array}\right)
\end{equation}
and 
\begin{equation}
\label{ }
\green^{HH}_R(\rv_t,\rv_t)=\frac{i\omega^2}{4\pi c^2}\left(\begin{array}{ccc}\int_0^\infty\frac{KdK}{2\gamma}(r^p-r^s\frac{\gamma^2}{k_0^2})e^{2i\gamma z_t} & 0 & 0 \\0 & \int_0^\infty\frac{KdK}{2\gamma}(r^p-r^s\frac{\gamma^2}{k_0^2})e^{2i\gamma z_t} & 0 \\0 & 0 & \int_0^\infty\frac{K^3dK}{\gamma k_0^2}r^se^{2i\gamma z_t}\end{array}\right)
\end{equation}

We suppose now 
that the separation distance is much smaller than the wavelength, so that the distance-dependent terms can be replaced by their asymptotic expressions.
In that case, $r^p=(\epsilon-1)/(\epsilon+1)$ and $r^s=(\epsilon-1)k_0^2/4K^2$. Integration over the parallel wavevector is easy since $\gamma^2=k_0^2-K^2\sim-K^2$ so that $\gamma=iK$. We recall that this is valid in a regime sometimes called "extreme near-field" \cite{Henkel:2000tr}, where the integrals are mostly due to their large $K$ contribution, which is not always the case \cite{Chapuis:2008kca} even in the near field. In this case, asymptotic expressions of the Green dyadics read

\begin{equation}
\label{ }
\green^{EE}_R(\rv_t,\rv_t)\approx\frac{\mu_0c^2}{32\pi z_t^3}\left(\begin{array}{ccc}\frac{\epsilon-1}{\epsilon+1} & 0 & 0 \\0 & \frac{\epsilon-1}{\epsilon+1} & 0 \\0 & 0 &2\frac{\epsilon-1}{\epsilon+1}\end{array}\right)
\end{equation}
and
\begin{equation}
\label{ }
\green^{HH}_R(\rv_t,\rv_t)=\frac{k_0^2}{8\pi z_t}\left(\begin{array}{ccc}\frac{1}{2}\left(\frac{\epsilon-1}{\epsilon+1}+\frac{\epsilon-1}{4}\right)& 0 & 0 \\0 & \frac{1}{2}\left(\frac{\epsilon-1}{\epsilon+1}+\frac{\epsilon-1}{4}\right) & 0 \\0 & 0 &\frac{\epsilon-1}{4}\end{array}\right)
\end{equation}
So that the effective polarizabilities take the form :
\begin{equation}
\label{alphaeffapprox}
\alpht^{E}\approx\left(\begin{array}{ccc}\frac{\alpha^E}{1-\frac{\alpha^E(\epsilon-1)}{32\epsilon_0\pi z_t^3(\epsilon+1)}} & 0 & 0 \\0 & \frac{\alpha^E}{1-\frac{\alpha^E(\epsilon-1)}{32\epsilon_0\pi z_t^3(\epsilon+1)}} & 0 \\0 & 0 &\frac{\alpha^E}{1-\frac{\alpha^E(\epsilon-1)}{16\epsilon_0\pi z_t^3(\epsilon+1)}}\end{array}\right)
\end{equation}
\begin{equation}
\label{ }
\alpht^{H}\approx\left(\begin{array}{ccc}\frac{\alpha^H}{1-\frac{\alpha^Hk_0^2}{16\pi z_t}\left(\frac{\epsilon-1}{\epsilon+1}+\frac{\epsilon-1}{4}\right)} & 0 & 0 \\0 & \frac{\alpha^H}{1-\frac{\alpha^Hk_0^2}{16\pi z_t}\left(\frac{\epsilon-1}{\epsilon+1}+\frac{\epsilon-1}{4}\right)} & 0 \\0 &0 &\frac{\alpha^H}{1-\frac{\alpha^Hk_0^2}{32\pi z_t}(\epsilon-1)}\end{array}\right)
\end{equation}
These expressions are the asymptotic approximations of the Green tensors. For the electric term, the obtained expression is also known as the electrostatic limit, which means that retardation is not taken into account. Note that the magnetic terms vanishes if retardation is discarded due to $\alpha^{H}$. 

\begin{figure}
\begin{center}
\includegraphics[width=12cm]{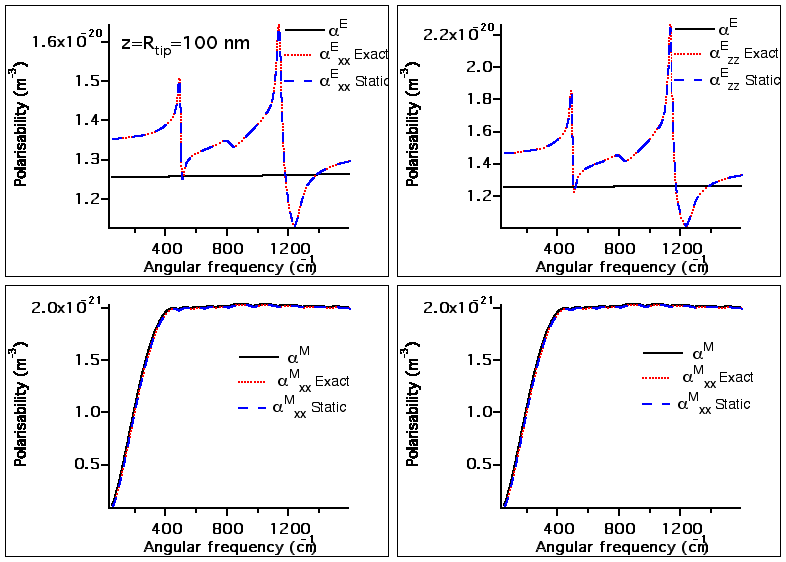}
\includegraphics[width=12cm]{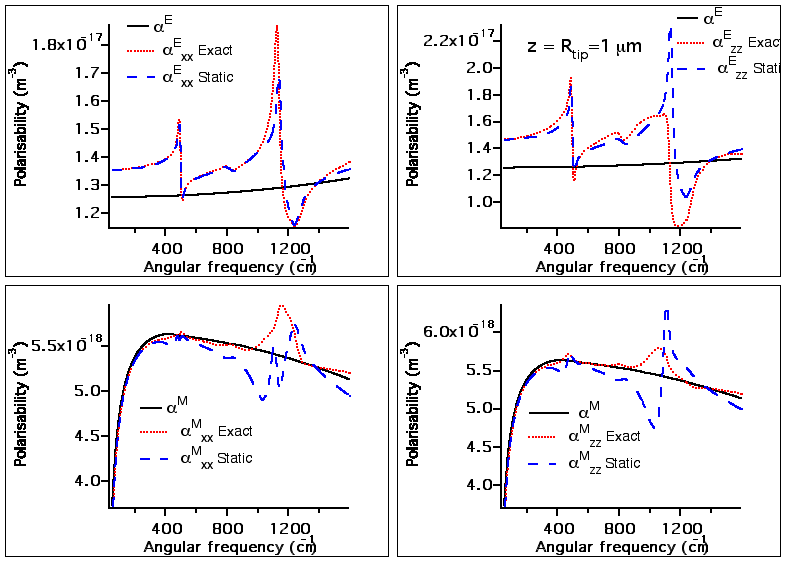}
\caption{Comparison between effective polarizabilities obtained with the exact theory developed here and obtained through electrostatic hypothesis. Four top figures : 100nm-radius W sphere situated at $z=100$ nm of a surface of SiO2. Four bottom figures : 1$ \mu$m-radius W sphere situated at $z$=1 $\mu$m of a surface of SiO2.}
\label{compkeil}
\end{center}
\end{figure}

The resulting expressions for the electric effective polarizabilities  are very close to previous work of Knoll and Keilmann\cite{Knoll:2000wm}. In Fig. \ref{compkeil}, the example of a tungsten spherical particle above SiO$_2$ shows that this approximation is very good for small particles at short distances. For larger particle, such as a 1$\mu$m particle located at a 1 $\mu$m distance, some deviation exists between the exact calculation and asymptotic expressions. This is due to retardation effects that enter into action and which are clearly not negligible even at subwavelegth distances as shown here.

Fig. \ref{compkeil} shows also that the effective magnetic polarisabilities and their approximations fit together at shortest distances but not anymore at larger ones. However, we note that for small particles, the correction at large distances (but still in the near field) is negligible due to the smallness of the polarizability. We underline also that the electrostatic approximation neglecting any magnetic term can be used only at extremely small distances.

\section{Application to SNOM detection}
\begin{figure}
\begin{center}
\includegraphics[width=12cm]{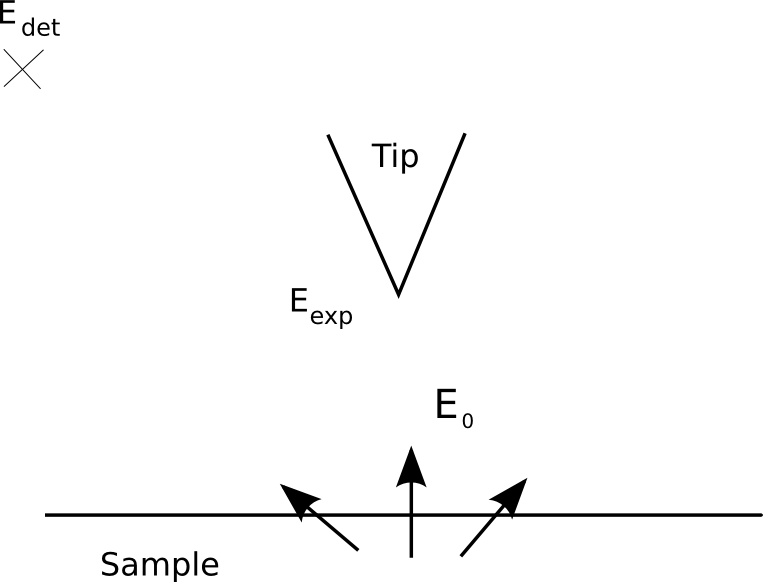}
\caption{Detection system scheme}
\label{detscheme}
\end{center}
\end{figure}

The spherical particles precedently studied model tips in SNOM experiments. The goal of this kind of experiments is to study the electromagnetic field close to a surface, the tip-surface distances being much smaller than the wavelength considered as a consequence (near field). The tip scatters the near-field electromagnetic radiation, which is then detected in the far-field by a detector located at a position $\rv_d$. The detection scheme is represented in Fig. \ref{detscheme}

\subsection{Signal at the detector}
The tip possesses electric and magnetic polarizabilities, so that the incoming electromagnetic field creates electric and a magnetic dipole moments at the tip position $\rv_t$. These dipoles radiate an EM field which is detected at the detector position, in the far field. Thus, the detected field is just the simple electromagnetic field radiated in far field by an electric and a magnetic dipoles taking into account the influence of the interface. The signal at the detector is calculated in the so-called far-field approximation, where the distance has to be large compared to square of scatterer size $\rho$ over the wavelength: $d>>\rho^2/\lambda$. The signal at the detector can be considered as a plane wave so that it reads
\begin{equation}
\label{signal1}
\left<S(\om)\right>=\frac{\epsilon_0c}{2}|\E^d(\om)|^2r^2d\Omega
\end{equation}

Let us introduce, the elementary vector of the tip-detector direction $\uv_d=(\rv_d-\rv_t)/|\rv_d-\rv_t|=\uv_{d\parallel}+\uv_{d\perp}$ where $\uv_{d\perp}$ is perpendicular to the interface. We define $\uv_{d}^-=\uv_{d\parallel}-\uv_{d\perp}$, $\shat_d=\uv_{d\parallel}\times\ev_z$, $\ppp_d=\uv_d\times\shat_d$ and $\ppm_d=-\uv_d^-\times\shat_d$. 
Let us also introduce four more tensors $\hperp(\uv_d)$, $\gperp(\uv_d)$, $\hperp^R(\uv_d)$ and $\gperp^R(\uv_d)$. 
\begin{eqnarray}
\hperp(\uv_d) & = & \shat_d\shat_d+\ppp_d\ppp_d \\
\gperp(\uv_d) & = & -\shat_d\ppp_d+\ppp_d\shat_d \\
\hperp^R(\uv_d) & = & (\shat_dr^s_d\shat_d+\ppp_dr^p_d\ppm_d)e^{i\phi} \\
\gperp^R(\uv_d) & = & (-\shat_dr^s_d\ppm_d+\ppp_dr^p_d\shat_d) e^{i\phi}
\end{eqnarray}
where $\phi$ denotes the phase difference between radiation going from the tip to the detector and the one going from the tip to the detector after one reflection on the surface.
Following classical expressions of field radiated by dipoles in the far field, the electromagnetic field radiated at the detector is
\begin{eqnarray}
\E^d & = & \frac{\mu_0\omega^2}{4\pi}\frac{e^{ikR}}{R}\left[\hperp(\uv_d)+\hperp^R(\uv_d)\right]\alpht^E\E^0(\rv_t)\nonumber \\
 & + & \frac{\mu_0\om^2}{4\pi c} \frac{e^{ikR}}{R}\left[\gperp(\uv_d)+\gperp^R(\uv_d)\right]\alpht^H\Hv^0(\rv_t)
\end{eqnarray}

 If we introduce now $\gamperp^E(\uv_d)=\hperp(\uv_d)+\hperp^R(\uv_d)$ and $\gamperp^H(\uv_d)=\gperp(\uv_d)+\gperp^R(\uv_d)$, we find, replacing  (\ref{signal1}) the expression of the power received at the detector in a solid angle $d\Omega$ around the detector direction:
\begin{equation}
\label{ }
\left<S\right>=\frac{\mu_0\om^4d\Omega}{32\pi^2c}\sum_{i,j,k}\left(\Gamma
^E_{ij}\alpha^E_{jj}E^0_j(\rv_t)+\frac{\Gamma^H_{ij}}{c}\alpha^H_{jj}H^0_j(\rv_t)\right)\left(\Gamma
^{E*}_{ik}\alpha^{E*}_{kk}E^{0*}_k(\rv_t)+\frac{\Gamma^{H*}_{ik}}{c}\alpha^{H*}_{kk}H^{0*}_k(\rv_t)\right)
\end{equation}
The final result is a combination of the different electromagnetic field components. These expressions are quadratic expressions of the electromagnetic field component and are related to the electromagnetic energy or the Poynting vector close to the interface.

\subsection{Expressions of the tensors $\hperp$ and $\gperp$.}
\begin{figure}[h]
\centering
\includegraphics[width=10cm]{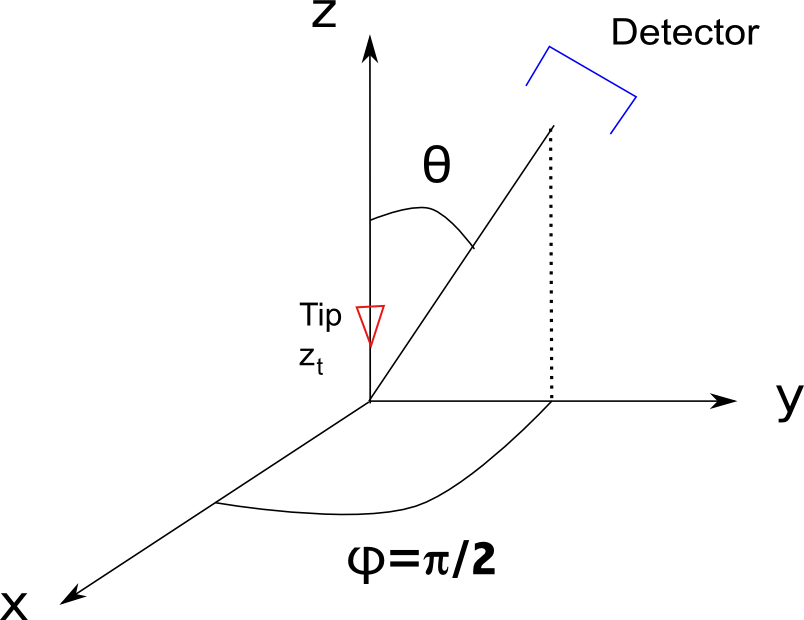}
\caption{Tip and detector position}
\label{tipos}
\end{figure}

We consider now a situation where the detector makes an angle $\theta$ with the vertical axis in the $z-y$ plane. Therefore, in spherical coordinates, the detector is at position $(r_d,\varphi,\theta)$ with $\varphi=\pi/2$ (Fig. \ref{tipos}). The tip is at position $z_t$ above the interface.

 At first order in $z_t$, the tensors reads
\begin{equation}
\label{ }
\hperp+\hperp^R=\left(\begin{array}{ccc}
1+r^s(\theta)e^{2ik_0z_t\cos\theta} & 0 & 0\\
0 & \cos^2\theta(1-r^p(\theta)e^{2ik_0z_t\cos\theta}) & -\sin\theta\cos\theta(1+r^p(\theta)e^{2ik_0z_t\cos\theta}) \\
0 & -\sin\theta\cos\theta(1-r^p(\theta)e^{2ik_0z_t\cos\theta}) & \sin^2\theta(1+r^p(\theta)e^{2ik_0z_t\cos\theta})
\end{array}\right)
\end{equation}
and
\begin{equation}
\label{ }
\gperp+\gperp^R=\left(\begin{array}{ccc}
0 & \cos\theta(1-r^s(\theta)e^{2ik_0z_t\cos\theta}) & -\sin\theta(1+r^s(\theta)e^{2ik_0z_t\cos\theta})\\
-\cos\theta(1+r^p(\theta)e^{2ik_0z_t\cos\theta}) & 0 & 0 \\
\sin\theta(1+r^p(\theta)e^{2ik_0z_t\cos\theta}) &0 & 0\end{array}\right)
\end{equation}

\section{Apertureless SNOM signal for various situations of interest}
\subsection{Signal generated by a surface plasmon excited with a quantum cascade laser}

We now calculate the signal detected by an apertureless SNOM above a material supporting surface plasmons. Typical materials are metals \cite{Raether:1988ty} such as gold and all materials for which the dielectric constant is smaller than -1. Surface plasmons are eigenmodes of a planar surface. They correspond to a pole of the plane interface reflection coefficient. They only exist for TM (or $p$) polarization.  If $z$ is the direction perpendicular to the interface, the plasmon magnetic field is given by
\begin{equation}
\label{ }
\Hv(\rv)=\left|\begin{array}{c}H_0 e^{i(Ky+\gamma_0z)} \\0 \\0\end{array}\right.
\end{equation}
whereas the electric field reads
\begin{equation}
\label{ }
\E(\rv)=\left|\begin{array}{c}0 \\-\frac{\gamma_0H_0}{\epsilon\omega}e^{i(Ky+\gamma_0z)} \\\frac{KH_0}{\epsilon\omega}e^{i(Ky+\gamma_0z)}\end{array}\right.
\end{equation}
If the detector makes an angle $\theta$ with the vertical axis and is situated in the $z-y$ plane, the signal at the detector is
\begin{eqnarray}
\left<S(\rv_d)\right> & = & \frac{\mu_0c\omega^4}{32\pi^2c^2}|H_0|^2e^{-2\Im(K)y}e^{-2\Im(\gamma_0)z} \label{sigdet} \\
 & \times & \left[\cos^2\theta|1-r^p(\theta)e^{2ik_0z_t\cos\theta}|^2|\alpha^E_{xx}|^2\frac{|\gamma_0|^2}{|\epsilon|^2\omega^2}+\sin^2\theta|1+r^p(\theta)|^2e^{2ik_0z_t\cos\theta}|\alpha^E_{zz}|^2\frac{|K|^2}{|\epsilon|^2\omega^2}\right.\nonumber\\
 &+&\frac{|1+r^p(\theta)e^{2ik_0z_t\cos\theta}|^2|\alpha^H_{xx}|^2}{c^2}\nonumber\\
 &+&\frac{2\sin\theta\cos\theta}{|\epsilon|^2\omega^2}\Re[(1-r^p(\theta)e^{2ik_0z_t\cos\theta})(1+r^{p*}(\theta)e^{-2ik_0z_t\cos\theta})\alpha^E_{xx}\alpha^{E*}_{xx}\gamma_0K^*]\nonumber\\
 &+&\frac{2\cos\theta}{c}\Re\left[(1-r^p(\theta)e^{2ik_0z_t\cos\theta})(1+r^{p*}(\theta)e^{2-ik_0z_t\cos\theta})\alpha^E_{xx}\alpha^{H*}_{xx}\frac{\gamma_0}{\epsilon\omega}\right]\nonumber\\
  &+&\left.\frac{2\sin\theta}{c}\Re\left[|1+r^p(\theta)e^{2ik_0z_t\cos\theta}|^2\alpha^E_{zz}\alpha^{H*}_{xx}\frac{K}{\epsilon\omega}\right]\right]\nonumber
\end{eqnarray}

\begin{figure}
\begin{center}
\includegraphics[width=10cm]{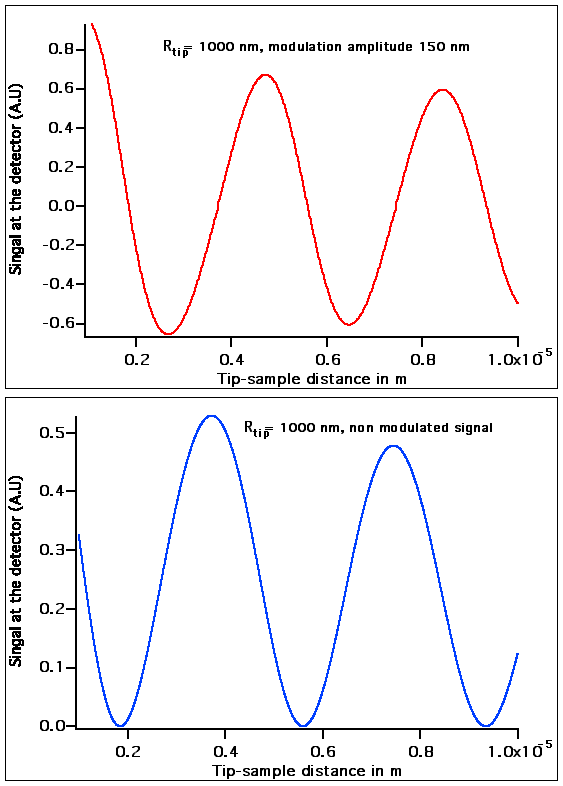}
\caption{Theoretical signal above a sample of gold on which plasmons have been excited at a wavelength $\lambda$ = 7.5 $\mu$m. Upper figure : signal detected by a 1$\mu$m-radius tungsten tip above a gold plane interface supporting a plasmon excited at $\lambda$ =7.5 $\mu$m. The tip oscillates with a 150 nm-amplitude. The amplitude of the first harmonic of the signal is plotted, corresponding to the signal detected at the tip oscillation frequency. Lower figure : same conditions as above except that the tip position is not modulated}
\label{Theoplasm}
\end{center}
\end{figure}

\begin{figure}
\begin{center}
\includegraphics[width=10cm]{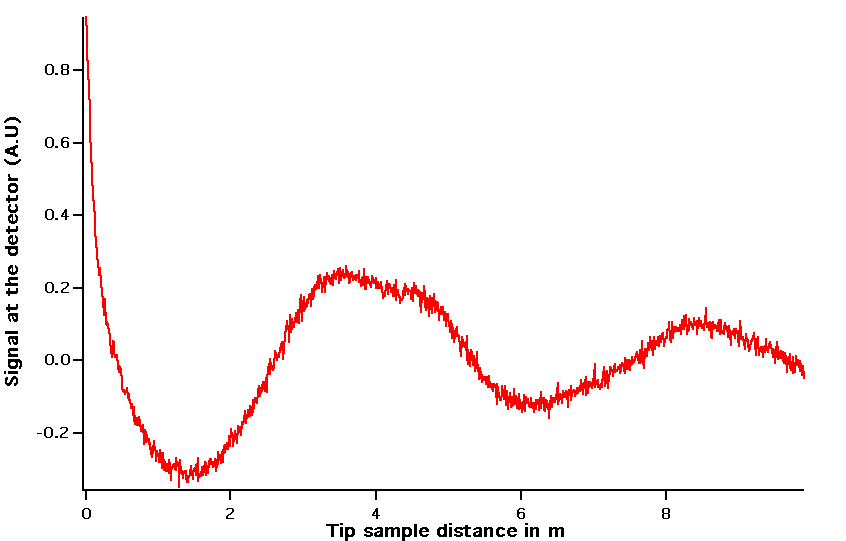}
\caption{SNOM signal above a gold sample supporting a plasmon excited by a quantum-cascade laser at 7.5 $\mu$m. Detecting tip is made of tungsten and oscilation ampitude is 150 nm.}
\label{Expplasm}
\end{center}
\end{figure}

In upper Fig. \ref{Theoplasm}, we plot the signal at the detector with the tip-sample distance in the case of a 1 $\mu$m-radius spherical dipole which simulates a ungsten tip oscillating in the vertical direction with an amplitude of 150 nm. These conditions are the ones used in SNOM experiments such as Ref \cite{DeWilde:2006kt}. We observe oscillations as the tip is retracted away from the surface. This is due to interferences between the signal scaterred by the tip and radiated directly to the detector and the signal scattered by the tip that undergoes a reflection before reaching the detector. The lower curve in Fig. \ref{Theoplasm} shows the same kind of signal in the case in which the tip is not modulated. The signal  behaviours obtained from an oscilating tip or an non-oscilating tip are similar. The oscillation period is around $\lambda_{plasmon}/2=3.75$ $\mu$m and the signal decreases with the tip-sample distance. We observe a similar behaviour when one probes the plasmonic signal above gold as seen in Fig. \ref{Expplasm}. Here the plasmon is excited by a quantum cascade laser at 7.5 $\mu$m. Besides the oscillations, the difference between the theoretical and the experimental curves may be due to different geometry of the signal collection, which involves a Cassegrain objective in the experiment, and to the fact that our model approximates the tip by a 1 µm radius tungsten sphere while it is in reality made scatterers distributed along a long cone with a curved apex.


\subsection{Signal due to thermal emission}
\subsubsection{Thermal fields above a sample}

 When the electromagnetic field above the surface is only the one due to thermal emission from the sample surface, several simplifications occurs. Cross correlations functions such as $\left<E^0_x(\rv_t)E^{0*}_y(\rv_t)\right>$, $\left<E^0_x(\rv_t)E^{0*}_z(\rv_t)\right>$, $\left<E^0_y(\rv_t)E^{0*}_z(\rv_t)\right>$, $\left<H^0_x(\rv_t)H^{0*}_y(\rv_t)\right>$, $\left<H^0_x(\rv_t)H^{0*}_z(\rv_t)\right>$, $\left<H^0_y(\rv_t)H^{0*}_z(\rv_t)\right>$, $\left<E^0_x(\rv_t)H^{0*}_z(\rv_t)\right>$, $\left<E^0_z(\rv_t)H^{0*}_x(\rv_t)\right>$, $\left<E^0_y(\rv_t)H^{0*}_z(\rv_t)\right>$ and $\left<E^0_z(\rv_t)H^{0*}_y(\rv_t)\right>$ are all equal to 0 due to the fact that thermal currents are decorrelated for different directions \cite{Joulain:2005ih}. Moreover, due to rotational symmetry around any axis in the $z$ direction, $\left<|E^0_x(\rv_t)|^2\right>=\left<|E^0_y(\rv_t)|^2\right>$ and $\left<|H^0_x(\rv_t)|^2\right>=\left<|H^0_y(\rv_t)|^2\right>$.
 
In these conditions, the signal at the detector in a direction making an angle $\theta$ with the $z$ axis reads
\begin{eqnarray}
\left<S(\rv_d)\right> & = & \frac{\mu_0\omega^4}{32\pi c}d\Omega\left\{\left(\cos^2\theta|1-r^p(\theta)e^{2ik_0z_t\cos\theta}|^2+|1+r^s(\theta)e^{2ik_0z_t\cos\theta}|^2\right)|\alpha^E_{xx}|^2|E^0_x(\rv_t)|^2\right.\nonumber\\
 & + & \sin^2\theta|1+r^pe^{2ik_0z_t\cos\theta}|^2 |\alpha^E_{zz}|^2|E^0_z(\rv_t)|^2+\frac{\sin^2\theta|1+r^se^{2ik_0z_t\cos\theta}|^2 }{c^2}|\alpha^E_{zz}|^2|H^0_z(\rv_t)|^2\nonumber \\
 &+&\frac{\cos^2\theta|1-r^s(\theta)e^{2ik_0z_t\cos\theta}|^2+|1+r^p(\theta)e^{2ik_0z_t\cos\theta}|^2}{c^2}|\alpha^H_{xx}|^2|H^0_x(\rv_t)|^2\\
 &+& 2\cos\theta\Re\left[\alpha^E_{xx}\alpha^{H*}_{xx}E^0_x(\rv_t)H^{0*}_y(\rv_t)\right.\nonumber\\
&\times& \left.\left.\left(\frac{(1+r^se^{2ik_0z_t\cos\theta})(1-r^{s*}e^{-2ik_0z_t\cos\theta})+(1-r^pe^{2ik_0z_t\cos\theta})(1+r^{p*}e^{-2ik_0z_t\cos\theta})}{c^2}\right)\right]\right\} \nonumber
\end{eqnarray}
Expressions of the electromagnetic field quadratic quantities due to thermal emission above a plane interface are well known in the literature\cite{Joulain:2005ih,Volokitin:2007el,Dorofeyev:2011bg} : 
\begin{eqnarray}
\left<|E^0_x(\rv_t)|^2\right> & = & \frac{\mu_0\om^2\Theta(\omega,T)}{2\pi^2c}\Im\left(i\int_0^\infty\frac{udu}{v}\left[1+r^se^{2ik_0vz}+v^2(1-r^pe^{2ik_0vz})\right]\right) \label{ex2}\\
\left<|E^0_z(\rv_t)|^2\right> & = & \frac{\mu_0\om^2\Theta(\omega,T)}{\pi^2c}\Im\left(i\int_0^\infty\frac{u^3du}{v}(1+r^pe^{2ik_0vz})\right)\\
\left<|H^0_x(\rv_t)|^2\right> & = &\frac{\epsilon_0\om^2\Theta(\omega,T)}{2\pi^2c}\Im\left(i\int_0^\infty\frac{udu}{v}\left[1+r^pe^{2ik_0vz}+v^2(1-r^se^{2ik_0vz})\right]\right) \\
\left<|H^0_z(\rv_t)|^2\right> & = & \frac{\epsilon_0\om^2\Theta(\omega,T)}{\pi^2c}\Im\left(i\int_0^\infty\frac{u^3du}{v}(1+r^se^{2ik_0vz})\right) \\
\left<E^0_x(\rv_t)H^{0*}_y(\rv_t)\right> &=&  \frac{\Theta(\omega,T)\omega^2}{4\pi^2c^2}\left[\int_0^\infty \frac{2uvdu}{|v|}\Re\frac{v}{|v|}(1+r^se^{2ik_0vz}-r^pe^{2ik_0vz})\right]\label{exhy}
\end{eqnarray}
where $v=\sqrt{1-u^2}$ and $\Theta(\om,T)=1/[\exp[\hbar\om/(k_bT)]-1]$ is the mean energy of an oscillator at angular frequency $\omega$ at thermal equilibrium.
As a consequence, if the substrate and the particle dipole materials are known, the signal at the detector coming from an emitted thermal electromagnetic field is known. If the signal is divided by the mean energy of an oscillator $\Theta(\om,T)$, it only depends on the particle polarizability, the tip-sample distance and the surface optical properties. If the polarizability is known, one can map a quantity which is specific of the surface and therefore make a surface spectroscopy.
Note however that expressions of the effective polarizability and of the thermal electromagnetic field are closely related. Indeed, both are depending of the system Green tensors taken at the tip position. As a consequence, if there is a resonance at a given frequency for the quadratic electromagnetic field expressions, it is likely that there will be also one in the effective polarizability spectra, at the same frequency. This constitutes a difficulty for the interpretation of the signal at the detector. Indeed, quadratic electromagnetic quantities are closely related to the electromagnetic LDOS (EM LDOS) \cite{Joulain:2005ih}. Relating the signal detector to the LDOS would be very interesting since it would give a way to detect this quantity as electronic tunneling microscopy does it for electronic LDOS \cite{Tersoff:1985wm}. Unfortunately, it is not possible to find a simple and universal relation between the two quantities. However, in some specific situations, when one term is leading, the EM LDOS and the detected signal can be proportional.

\subsubsection{Probing a polar material}

\paragraph{Signal simulation}
\begin{figure}
\begin{center}
\includegraphics[width=12cm]{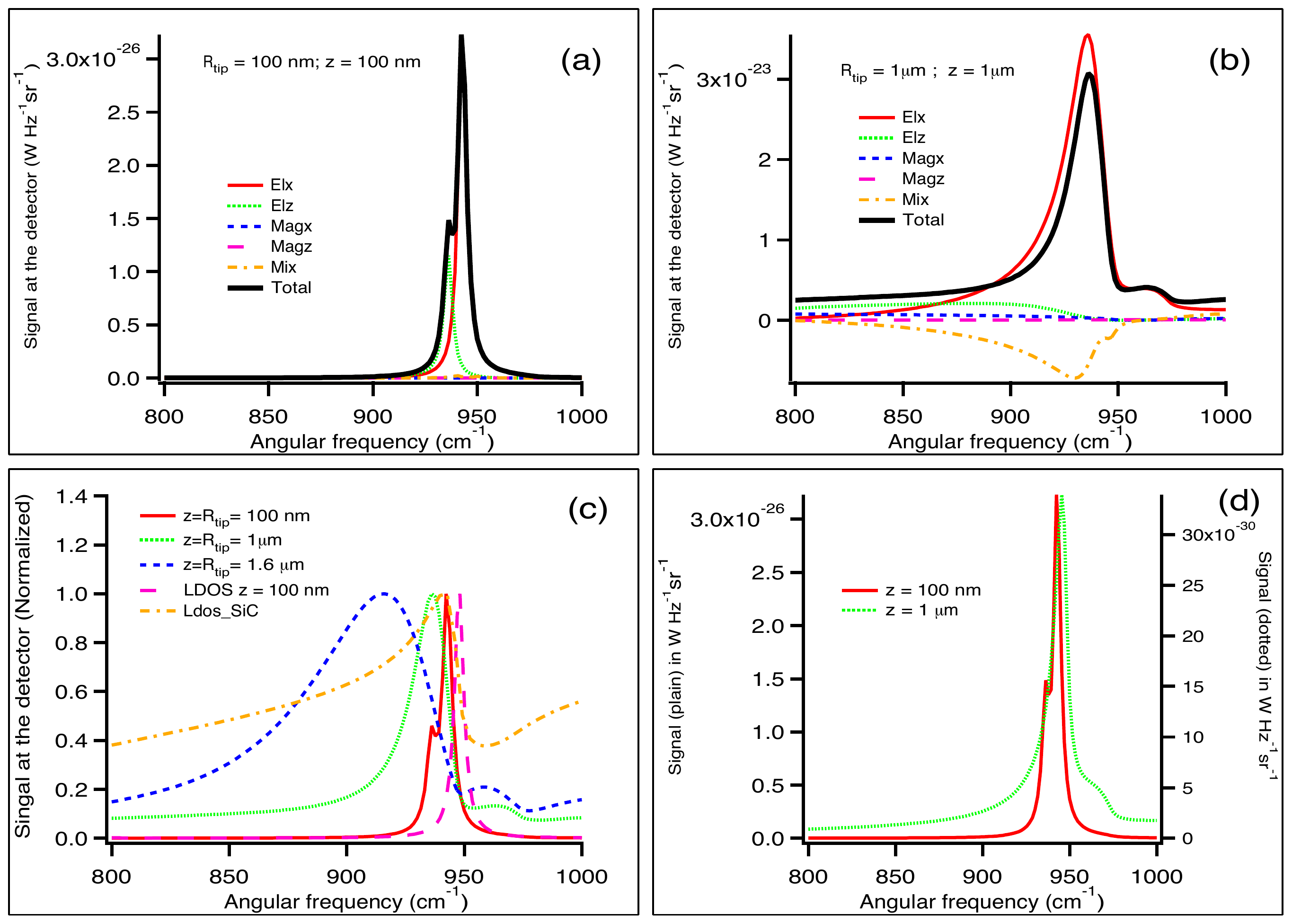}
\caption{Signal spectrum at the detector. The tip is constituted of a W spherical particle situated above SiC. The signal is detected in a direction making an angle of 30$^0$ with the vertical (a) Total signal and different contributions for a 100nm radius tip at $z=$100nm above the surface. (b) Total signal and different contributions for a 1 $\mu$m radius tip at $z=$1 $\mu$m above the surface. (c) Signal spectrum for $z=R_{tip}=100$ nm, $z=R_{tip}=1$ $\mu$m and $z=R_{tip}=1.6$ $\mu$m. Comparison with LDOS (d) Signal spectrum for $R_{tip}=100$ nm and two different distances to the interface $z=100$ nm and $z=1.6$ $\mu$m.}
\label{sigdetWSiC}
\end{center}
\end{figure}

We now study the spectrum of the signal detected by a probe above SiC. As we have seen in the previous section, this signal strongly depends on the effective polarizability and therefore on the tip size and the tip distance. To illustrate this, we show in Fig. \ref{sigdetWSiC} the spectral signal detected by a 100nm tungsten probe above SiC and by a 1$\mu$m-radius probe. It is observed that the emission peak is rather narrow for the small tip. This peak is also situated around the phonon polariton resonance frequency that is 943 cm$^{-1}$ for SiC.
We also note that for a small tungsten tip, the signal above SiC is dominated at short distance by electrical terms. We see that both parallel and perpendicular terms have a significant contribution. These contributions are slightly shifted with respect to each other: Effective parallel and perpendicular resonances are indeed different with a factor of 2 (Eq. \ref{alphaeffapprox}) in the right terms of the denominator asymptotic expressions, leading to a different frequencies for the resonant peak

For a 1 $\mu$m tip situated at 1 $\mu$m from the interface, we note that the main contribution comes from the parallel electric term. Magnetic terms are also more important here than for small tip due to the fact  that the magnetic polarizability is larger. These magnetic terms do not contribute a lot because magnetic energy density is lower than electric energy near the polariton resonance for polar materials. However, the mixed term, which involves both magnetic and electric terms is here the second contribution to the signal and is not negligible. 
Fig. \ref{sigdetWSiC} (c) shows various signal spectra for tips that would be in contact with the surface. They are compared with the theoretical spectrum of the LDOS at 100 $nm$. A 100nm-radius tip gives a signal similar to the electromagnetic LDOS: A well-defined peak appears around the polariton frequency. The peak width is very close to the LDOS one but the peak position is nevertheless slightly shifted. When the tip size increases, the peak 
tends to shift and to broaden. 
Recent SNOM experiments based on the measurement of the near-field thermal emission using a tungsten tip seem to confirm this shifiting and broadening suggesting that the approximation of the tip by a simple spherical dipole which we make in our model is valid to some extent.
Note also that this broadening, although less pronounced than for a large tip, also occurs when a small tip is retracted from the surface as it can be seen in Fig. \ref{sigdetWSiC} (d). 

\paragraph{Relation between the detected signal and the LDOS}

As shown in Fig. \ref{sigdetWSiC}, the signal calculated with a small tip is very similar to the LDOS spectrum around the polariton resonance. At short distance, the signal is indeed mainly dominated by the parallel electric contribution. This means that the signal mainly depends on $\alpha^E_{xx}$ and on $|E^0_x|^2$. In the case of a thermal signal, this last quantity is representative of the electrical energy density and of the electromagnetic LDOS. Therefore, a SNOM experiment detecting thermal near field will measure the product of one of the effective polarizability terms by a partial contribution to the EM LDOS. If the effective polarizability has a flat response i.e. its does not vary with the frequency (or only weakly), one can say that the signal at the detector is proportional to the EM LDOS. This is not the case in the present situation where polarizability can be increased by a factor of 3 to 5 around the resonance: The resulting signal is the product of two peaks situated approximately at the same frequency. As the value of the EM LDOS in the peak spectral band is abut 100 to 1000 times its value outside this band, its multiplication by the effective polarizability will give a peak that is not exactly the EM LDOS but which is representative of the LDOS.    

We now consider larger tips of micrometer size radii. The signal at the detector scattered by such tips has a broader spectrum and the frequency corresponding to the emission spectrum is shifted to lower frequencies. 
When the tip size increases, polarizability is shifted to lower frequency and broadened. A larger distances, EM LDOS is also broadened and shifted but differently than the polarizability. Frequency shift is indeed less important for EM LDOS. The resulting signal is now the product of two peaks which are rather similar in terms of broadness but with their maxima separated. The resulting signal is a peak situated between the polarizability and LDOS peaks. With these conditions, the signal at the detector is a peak related to the EM LDOS but which is not stricly-speaking the EM LDOS.

\subsubsection{Probing a metallic sample}

\begin{figure}
\begin{center}
\includegraphics[width=12cm]{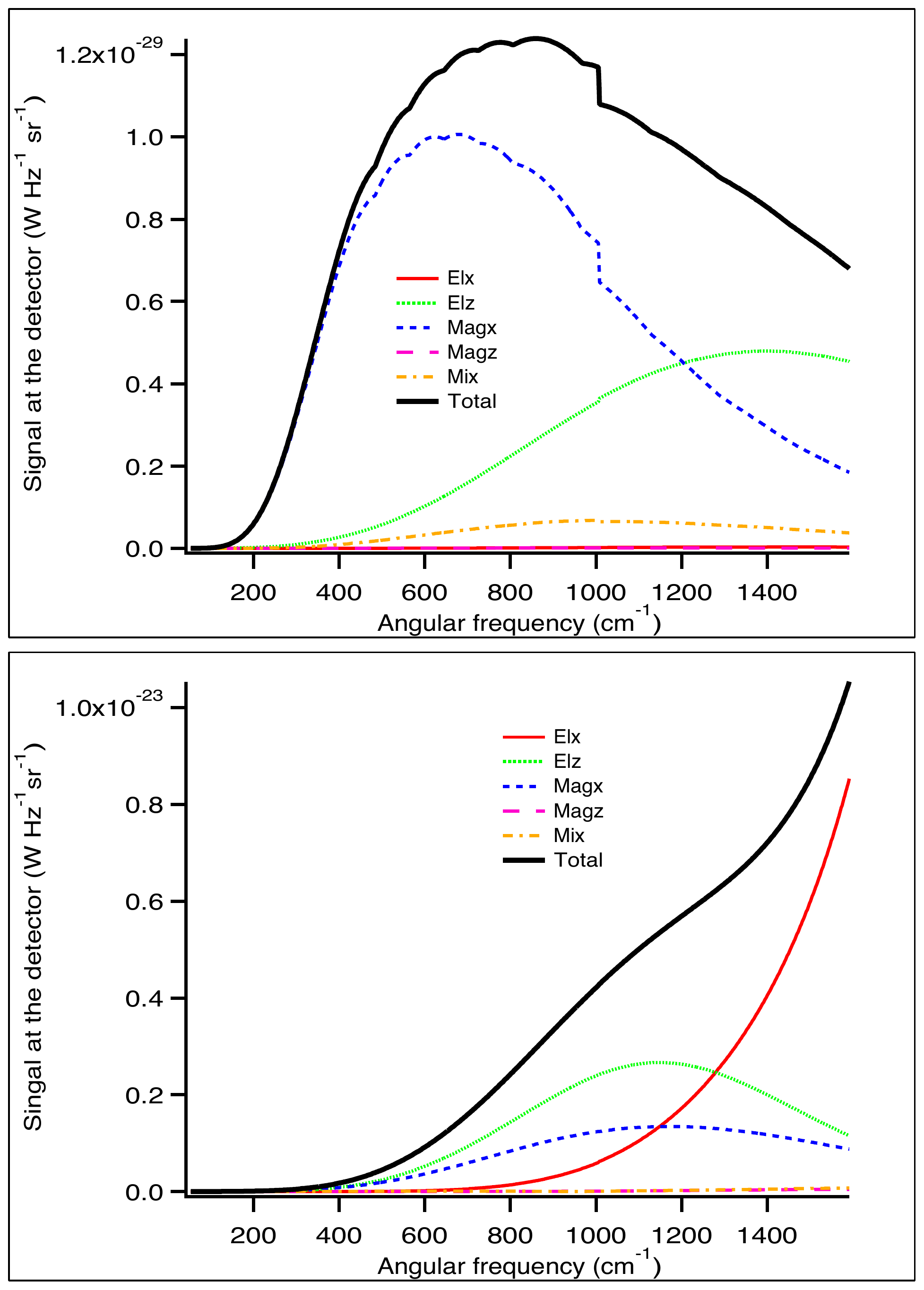}
\caption{Detected signal spectrum. The tip is constituted of a W spherical particle situated above Au. The signal is detected in a direction making an angle of 30$^0$ with the vertical (upper) Total signal and different contributions for a 100nm radius tip at $z=$100nm above surface. (Lower) Total signal and different contributions for a 1 $\mu$m radius tip at $z=$1 $\mu$m above surface.}
\label{sigdetWAu}
\end{center}
\end{figure}

We now consider the example of a tip situated above a metal surface. Fig. \ref{sigdetWAu} shows the signal scattered by a tungsten particle above a gold surface heated at 300 K. For a small particle, the contribution to the signal is dominated by the magnetic term below 1000 cm$^{-1}$. The main reason for that is the fact that the magnetic energy is more important than the electrical energy in this part of the spectrum where Au is highly reflecting. As seen in Fig. \ref{aleff100100}, the magnetic polarizability has a flat response for small tips above 400 cm$^{-1}$. Then, between 400 and 1000 cm$^{-1}$ the signal at the detector is proportional to parallel magnetic energy which is close to total electromagnetic energy. Thus, the signal detected in this spectral range is close to the EM LDOS. On the contrary, above 1000 cm$^{-1}$, the signal is dominated by the perpendicular electric contribution but the parallel magnetic and the mixed term are of the same order of magnitude even if smaller. For such situation, it is not correct to state that the signal is proportional to the EM LDOS although it is related to it. 
For a large tip, there is no spectral range where a term clearly dominates the contribution to the signal (except at high frequency where the validity of the dipole approximation becomes questionable).
The detected signal cannot be considered as simply proportional to the EM LDOS.

\section{Thermal emission of a nanoparticle}
We now consider a small dipolar particle heated at temperature $T$. This particle radiates an electromagnetic field in all space. This field is a thermal field which is here detected in far-field by a detector. The signal at the detector has a contribution coming directly from the particle in straight line or after one reflection at the interface. The signal at the detector has the following expression:

\begin{eqnarray}
\label{ }
\left<S(\omega)\right>&=&\frac{d\Omega}{32\pi^3}\Theta(\om,T)\frac{\om^3}{c^3}\left[\Im[\alpha^E_{xx}]\left(|1+r^s(\theta)e^{2ik_0z_t\cos\theta}|^2+\cos^2\theta|1-r^p(\theta)e^{2ik_0z_t\cos\theta}|^2\right)\right.\nonumber\\
&+&\left.\Im[\alpha^E_{zz}]|1+r^p(\theta)e^{2ik_0z_t\cos\theta}|^2\sin^2\theta\right]
\end{eqnarray}
In this expression, the spectral dependance comes mainly from the effective polarizability and also from the far field reflection coefficients. One thus expects that the detected spectral signal detected will follow polarizability variations that we have described in the preceding sections.

\begin{figure}
\begin{center}
\includegraphics[width=12cm]{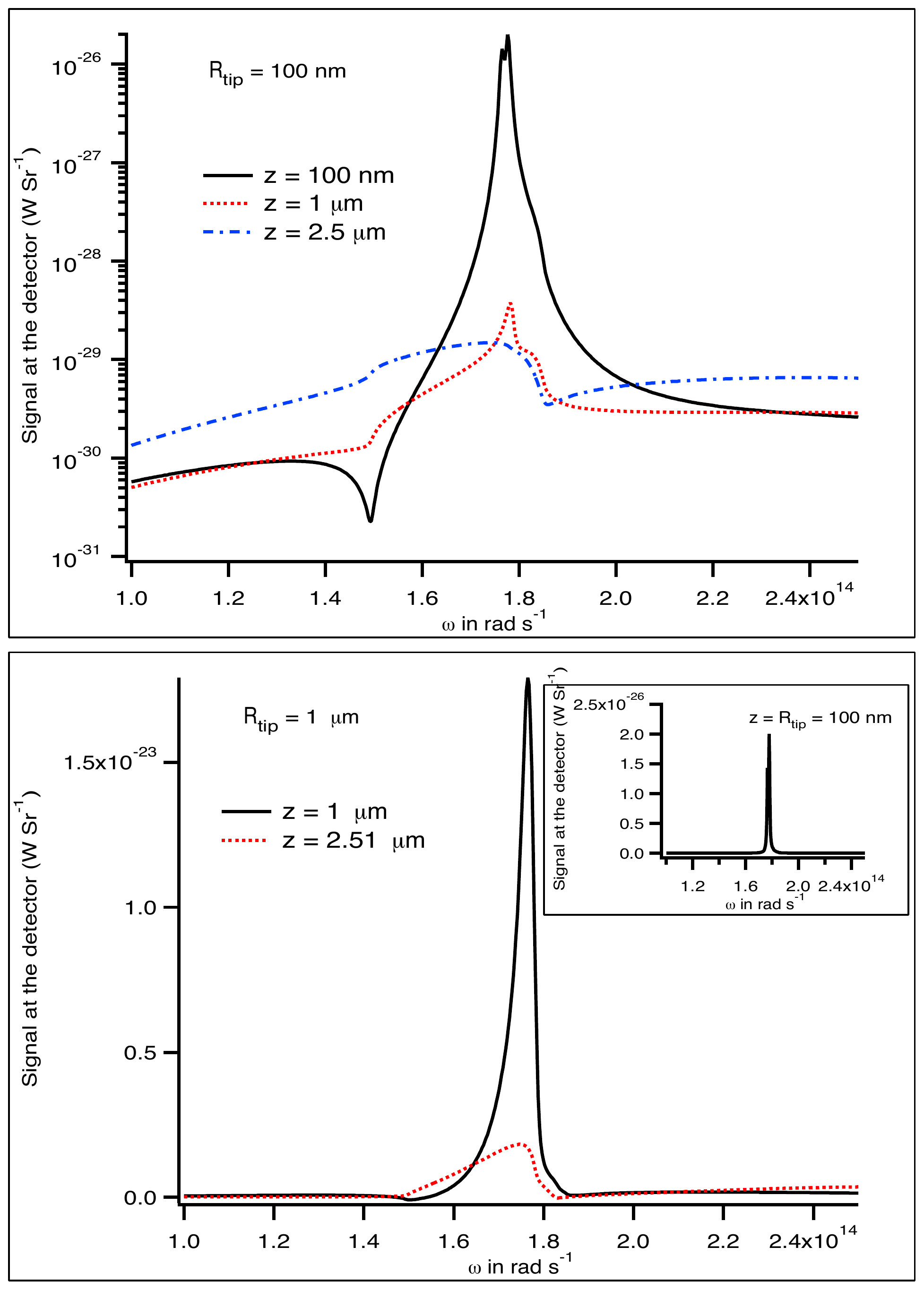}
\caption{Signal at the detector which direction makes an angle $\theta = 45 ^0C$ with the vertical axis. The tip is a spherical tip of tungsten heated at $T = 300$ K and situated at various particle-SiC interface. Upper figure : signal in log-scale for $R_{tip}$= 100 nm and for three different tip-sample distance. Lower figure : signal in linear scale for $R_{tip}$= 1 $\mu$m and for separation distance equal to 1 $\mu$m and 2.51 $\mu$m. Inset : signal for $z=R_{tip}$= 100 nm.}
\label{Chff}
\end{center}
\end{figure}

In Fig. \ref{Chff}, we plot the signal detected in far field by a detector in the 45$^0$ direction from the vertical  direction, in the case of a 100 nm-radius  tungsten spherical particle heated at 300 K and situated at different distances from the interface separating vacuum from SiC. When the particle-interface distance is 100 nm, the signal is peaked around SiC plasmon resonance, as is 100 nm-radius tungsten spherical particle polarizability. When the particle is retracted from the interface, the signal is reduced and broadens. Above a certain distance, the effective polarizability is nothing but the polarizability in vacuum. For tungsten, this polarizability has a rather flat dependance with the frequency. The signal spectral behaviour mainly comes from spectral variations of the reflection coefficient. One notes that the signal is more important in the frequency range where SiC is known to be highly reflective i.e. between 1.6$\times$ 10$^{14}$ rad s$^{-1}$ and 1.8$\times$ 10$^{14}$ rad s$^{-1}$ .
A similar behaviour is observed for a micrometric tip except that the signal is broader at the minimum distance. This can easily be explained by inspecting effective polarizability of a 1 $\mu$m-radius spherical tungsten tip situated at 1 $\mu$m of a SiC interface (tip in contact) where the noticed broadening is indeed observed. 

Note that for such situation, the detected signal can exhibit a peak very similar to the one observed in the energy density spectrum that is in the electromagnetic LDOS. However, this peak is the signature of the effective polarizability which exhibits a resonance at frequencies close to surface resonance. Even if such experiments do not probe EM LDOS, it can probe surface resonances if the tip is sufficiently close to the interface. It can then also be seen as a surface spectroscopy method.

\section{Radiative Cooling of a dipolar particle}

When a particle is heated, one can calculate the rate at which it will exchange energy with the outside. The spectral power lost by a particle at temperature $T$ represented by its dipole when vacuum at null temperature  is\cite{Joulain:2008tp}   
\begin{equation}
\label{ }
P(\om)d\om=\frac{\om^3}{\pi^2c^3}Im[\alpha(\om)]
\Theta(\om,T)d\om
\end{equation}
It has been shown in the past that when the particle is close to a surface it exchanges with it with a dependence of the polarizability and of the electromagnetic energy density at the particle position. This phenomenon is very similar to the Purcell effect when an atom or a molecule has its spontaneous emission rate that is modified when approached close to a surface. The cooling rate depends on the electromagnetic LDOS\cite{BenAbdallah:2011tz} in a similar way as the spontaneous emission rate depends on the EM LDOS. 
The cooling rate formula given in the paper of Mulet et al.\cite{Mulet:2001kp} should be corrected at close distances. Indeed, the interaction of the particle with the surface should be taken into account through the effective polarizability. The formula of the heat exchanged between the particle and the sample reads
\begin{equation}
\label{ }
P(\om)d\om=\frac{2}{\pi}\frac{\om^2}{c^2}\Theta(\om,T)\sum_{i=x,y,z}\Im(\alpha^E_{ii})\Im\left[ G^{EE}_{ii}(\rv_p,\rv_p)\right]d\om
\end{equation}

\begin{figure}
\begin{center}
\includegraphics[width=12cm]{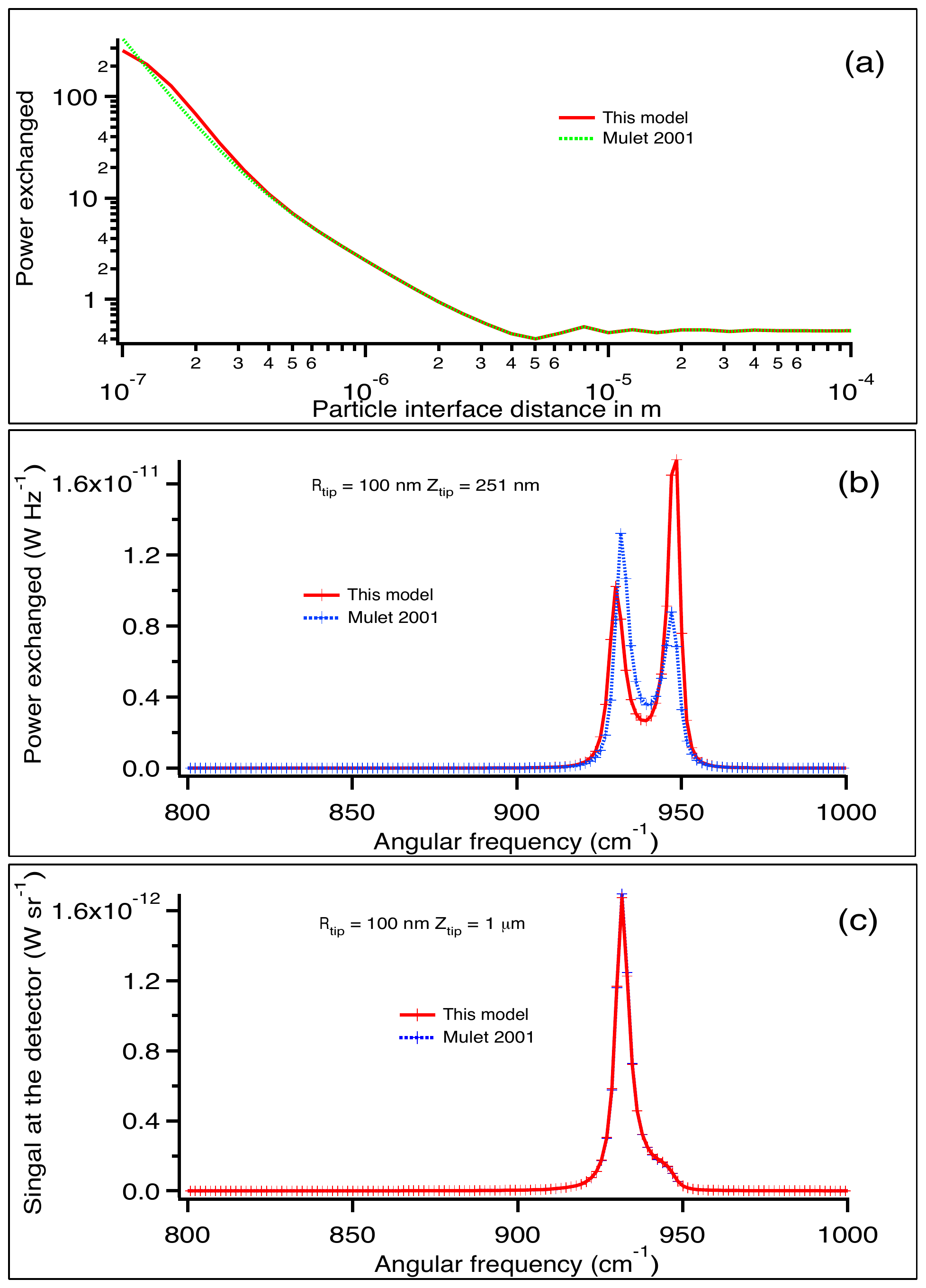}
\caption{(a) Heat transfer between a 100nm radius SiC particle at 300 K above a SIC substrate at null temperature versus distance. (b) Power exchanged spectrum between a 100 nm radius SiC particle above a SiC substrate at zero temperature situated at 251 nm below the particle. (c) Power exchanged spectrum between a 100 nm radius SiC particle above a SiC substrate at zero temperature situated at 1 $\mu$m below the particle.  }
\label{refr}
\end{center}
\end{figure}

In Fig. \ref{refr}, we plot the heat exchanged between a SiC particle and a SiC substrate at 300 K as a function of the tip-sample distance. We also show the spectrum of the exchanged flux. We compare, in each plot, the expression presented above and the one obtained by Mulet \cite{Mulet:2001kp}. In the plot of the exchanged power as a function of the particle-surface distance, we observe that both expressions give similar results down to distances as low as 400 nm. 
At large distances, corrections to the polarizability are small 
. The exchanged heat comes mainly from the spectral region where a resonance occurs in the SiC particule, around 935 cm$^{-1}$ i.e. where $\epsilon=-2$. 
When one enters in the near field, two phenomena occur : electromagnetic field is more-strongly dominated by evanescent contributions, in particular those coming from the phonon polariton resonance occuring around the frequency where $\epsilon$=-1. Moreover, when the distance between the particle and the interface is reduced, corrections to the polarizability appear. In the present case, the particle resonance peak is weakened, which explains the fact that at 100 nm, the exact expression of the transfer gives a lower contribution than the one used by Mulet et al. \cite{Mulet:2001kp}.

\section{Conclusion}

We have shown how the interaction between a probing SNOM tip and the surface it scans modifies infrared electromagnetic near fields. We have shown that the particle dipole polarizabilities can be modified to account for this interaction, and we have added the effect of retardation to the one of the image dipole. We have analysed the relation between the signal detected in the far-field by an apertureless SNOM and the excited fields, either due to external excitation of a plasmon or due to thermal emission. The technique performs a local spectroscopy of the surface and for some cases the signal spectrum can even be  proportional to the EM LDOS, a fundamental quantity. We have also corrected the near-field radiative heat transfer formula for cases where there is a strong particle-sample interaction. In the future, this theory should be extended to tips with more elongated shapes \cite{Huth:2010fg} such as conical ones, often used in experiment, in order to be applied to more realistic situations. 


\begin{acknowledgements}
The authors thanks Jean-Jacques Greffet for fruitful discussions. K.Joulain, Y. De Wilde and P. Ben-Abdallah thank ANR contracts NANOFTIR 07-NANO-039 and SOURCES-TPV  BLAN-0928-01 for financial supports. Y. De Wilde and A. Babuty thank the team of R. Colombelli and A. Bousseksou for fruitfull collaboration on plasmonics, and the CNANO IdF for PhD grant financial support with the project PSTST.  
\end{acknowledgements}


\begin{thebibliography}{0}%
\makeatletter
\providecommand \@ifxundefined [1]{%
 \@ifx{#1\undefined}
}%
\providecommand \@ifnum [1]{%
 \ifnum #1\expandafter \@firstoftwo
 \else \expandafter \@secondoftwo
 \fi
}%
\providecommand \@ifx [1]{%
 \ifx #1\expandafter \@firstoftwo
 \else \expandafter \@secondoftwo
 \fi
}%
\providecommand \natexlab [1]{#1}%
\providecommand \enquote  [1]{``#1''}%
\providecommand \bibnamefont  [1]{#1}%
\providecommand \bibfnamefont [1]{#1}%
\providecommand \citenamefont [1]{#1}%
\providecommand \href@noop [0]{\@secondoftwo}%
\providecommand \href [0]{\begingroup \@sanitize@url \@href}%
\providecommand \@href[1]{\@@startlink{#1}\@@href}%
\providecommand \@@href[1]{\endgroup#1\@@endlink}%
\providecommand \@sanitize@url [0]{\catcode `\\12\catcode `\$12\catcode
  `\&12\catcode `\#12\catcode `\^12\catcode `\_12\catcode `\%12\relax}%
\providecommand \@@startlink[1]{}%
\providecommand \@@endlink[0]{}%
\providecommand \url  [0]{\begingroup\@sanitize@url \@url }%
\providecommand \@url [1]{\endgroup\@href {#1}{\urlprefix }}%
\providecommand \urlprefix  [0]{URL }%
\providecommand \Eprint [0]{\href }%
\@ifxundefined \urlstyle {%
  \providecommand \doi  [0]{\begingroup \@sanitize@url \@doi}%
  \providecommand \@doi [1]{\endgroup \@@startlink {\doibase
  #1}doi:\discretionary {}{}{}#1\@@endlink }%
}{%
  \providecommand \doi  [0]{doi:\discretionary{}{}{}\begingroup
  \urlstyle{rm}\Url }%
}%
\providecommand \doibase [0]{http://dx.doi.org/}%
\providecommand \Doi [0]{\begingroup \@sanitize@url \@Doi }%
\providecommand \@Doi  [1]{\endgroup\@@startlink{\doibase#1}\@@Doi}%
\providecommand \@@Doi [1]{#1\@@endlink}%
\providecommand \selectlanguage [0]{\@gobble}%
\providecommand \bibinfo  [0]{\@secondoftwo}%
\providecommand \bibfield  [0]{\@secondoftwo}%
\providecommand \translation [1]{[#1]}%
\providecommand \BibitemOpen [0]{}%
\providecommand \bibitemStop [0]{}%
\providecommand \bibitemNoStop [0]{.\EOS\space}%
\providecommand \EOS [0]{\spacefactor3000\relax}%
\providecommand \BibitemShut  [1]{\csname bibitem#1\endcsname}%
\end{thebibliography}%


\begin{thebibliography}{10}


\bibitem{Rytov:1989ur}
S.M. Rytov, Y.A. Kravtsov, and V.T. Tatarskii.
\newblock {\em {Principle of Statistical Radiophysics 3}}, volume~3 of {\em
  Elements of Radiation Fields}.
\newblock Springer Verlag, 1989.

\bibitem{Joulain:2005ih}
Karl Joulain, Jean-Philippe Mulet, Fran{\c c}ois Marquier, R{\'e}mi Carminati,
  and Jean-Jacques Greffet.
\newblock {Surface electromagnetic waves thermally excited: Radiative heat
  transfer, coherence properties and Casimir forces revisited in the near
  field}.
\newblock {\em Surface Science Reports}, 57(3-4):59--112, May 2005.

\bibitem{Volokitin:2007el}
A~Volokitin and B~Persson.
\newblock {Near-field radiative heat transfer and noncontact friction}.
\newblock {\em Reviews of Modern Physics}, 79(4):1291--1329, October 2007.

\bibitem{Dorofeyev:2011bg}
I~A Dorofeyev and E~A Vinogradov.
\newblock {Fluctuating electromagnetic fields of solids}.
\newblock {\em Physics Reports}, 504(2-4):75--143, July 2011.

\bibitem{Polder:1971uu}
D~Polder and M.~van Hove.
\newblock {Theory of Radiative Heat Transfer between Closely Spaced Bodies }.
\newblock {\em Physical Review B}, 4:3303--3314, November 1971.

\bibitem{BenAbdallah:2010hp}
Philippe Ben-Abdallah and Karl Joulain.
\newblock {Fundamental limits for noncontact transfers between two bodies}.
\newblock {\em Physical Review B}, 82(12), September 2010.

\bibitem{Shchegrov:2000td}
A.V. Shchegrov, K~Joulain, R~Carminati, and J~J Greffet.
\newblock {Near-field Spectral Effects due to Electromagnetic Surface
  Excitations}.
\newblock {\em Physical Review Letters}, 85:1548--1551, August 2000.

\bibitem{Henkel:2000tr}
C~Henkel, K~Joulain, R~Carminati, and J~J Greffet.
\newblock {Spatial Coherence of thermal near-fields}.
\newblock {\em Optics Communications}, 186:57--67, December 2000.

\bibitem{Greffet:2002ur}
Jean-Jacques Greffet, R{\'e}mi Carminati, Karl Joulain, Jean-Philippe Mulet,
  St{\'e}phane Mainguy, and Yong Chen.
\newblock {Coherent emission of light by thermal sources}.
\newblock {\em Nature}, 416:61--63, March 2002.

\bibitem{Lee:2006cj}
B.J. Lee and Z~M Zhang.
\newblock {Design and fabrication of planar multilayer structures with coherent
  thermal emission characteristics}.
\newblock {\em Journal of Applied Physics}, 100:063529, September 2006.

\bibitem{Biener:2008cj}
Gabriel Biener, Nir Dahan, Vladimir Kleiner, and Erez Hasman.
\newblock {Highly coherent thermal emission obtained by plasmonic bandgap
  structures}.
\newblock {\em Applied Physics Letters}, 92:081913, February 2008.

\bibitem{Kittel:2005fr}
Achim Kittel, Wolfgang M{\"u}ller-Hirsch, J{\"u}rgen Parisi, Svend-Age Biehs,
  Daniel Reddig, and Martin Holthaus.
\newblock {Near-Field Heat Transfer in a Scanning Thermal Microscope}.
\newblock {\em Physical Review Letters}, 95(22), November 2005.

\bibitem{Narayanaswamy:2008gj}
Arvind Narayanaswamy, Sheng Shen, and Gang Chen.
\newblock {Near-field radiative heat transfer between a sphere and a
  substrate}.
\newblock {\em Physical Review B}, 78(11), September 2008.

\bibitem{Rousseau:2009es}
Emmanuel Rousseau, Alessandro Siria, Guillaume Jourdan, Sebastian Volz, Fabio
  Comin, Jo{\"e}l Chevrier, and Jean-Jacques Greffet.
\newblock {Radiative heat transfer at the nanoscale}.
\newblock {\em Nature Photonics}, 3:514--517, August 2009.

\bibitem{Ottens:2011kh}
R~Ottens, V~Quetschke, Stacy Wise, A~Alemi, R~Lundock, G~Mueller, D~Reitze,
  D~Tanner, and B~Whiting.
\newblock {Near-Field Radiative Heat Transfer between Macroscopic Planar
  Surfaces}.
\newblock {\em Physical Review Letters}, 107(1), June 2011.

\bibitem{Kittel:2008bc}
A~Kittel, U~F Wischnath, J~Welker, O~Huth, F~R{\"u}ting, and S~A Biehs.
\newblock {Near-field thermal imaging of nanostructured surfaces}.
\newblock {\em Applied Physics Letters}, 93(19):193109, 2008.

\bibitem{Wischnath:2008hp}
Uli~F Wischnath, Joachim Welker, Marco Munzel, and Achim Kittel.
\newblock {The near-field scanning thermal microscope}.
\newblock {\em Review of Scientific Instruments}, 79(7):073708, 2008.

\bibitem{DeWilde:2006kt}
Yannick De~Wilde, Florian Formanek, R{\'e}mi Carminati, Boris Gralak,
  Paul-Arthur Lemoine, Karl Joulain, Jean-Philippe Mulet, Yong Chen, and
  Jean-Jacques Greffet.
\newblock {Thermal radiation scanning tunnelling microscopy}.
\newblock {\em Nature}, 444(7120):740--743, December 2006.

\bibitem{Kajihara:2010fo}
Yusuke Kajihara, Keishi Kosaka, and Susumu Komiyama.
\newblock {A sensitive near-field microscope for thermal radiation}.
\newblock {\em Review of Scientific Instruments}, 81(3):033706, 2010.

\bibitem{Kajihara:2011uu}
Yusuke Kajihara, Keishi Kosaka, and Susumu Komiyama.
\newblock {Thermally excited near-field radiation and far-field interference}.
\newblock {\em Optics Express}, 19:7695--7704, April 2011.

\bibitem{Babuty:dYmzb9en}
Arthur Babuty, Karl Joulain, Pierre-Olivier Chapuis, Yannick De~Wilde, and
  Jean-Jacques Greffet.
\newblock {Probing near-field thermal spectra}.
\newblock {\em Submitted}, February 2012.

\bibitem{GarciadeAbajo:2007eb}
F~J Garc{\'\i}a~de Abajo.
\newblock {Colloquium: Light scattering by particle and hole arrays}.
\newblock {\em Reviews of Modern Physics}, 79(4):1267--1290, October 2007.

\bibitem{Intravaia:2010gp}
F~Intravaia, C~Henkel, and M~Antezza.
\newblock {Fluctuation-induced forces between atoms and surfaces: the
  Casimir-Polder interaction}.
\newblock In D~Dalvit, P~Milonni, D~Roberts, and F~da~Rosa, editors, {\em
  Lecture Notes in Physics}, pages 345--391. arXiv.org, October 2010.

\bibitem{BenAbdallah:2011be}
Philippe Ben-Abdallah, Svend-Age Biehs, and Karl Joulain.
\newblock {Many-Body Radiative Heat Transfer Theory}.
\newblock {\em Physical Review Letters}, 107(11), September 2011.

\bibitem{Castanie:2011ue}
{\'E}tienne Castani{\'e}, R{\'e}mi Vincent, Romain Pierrat, and R{\'e}mi
  Carminati.
\newblock {Absorption by an optical dipole antenna in a structured
  environment}.
\newblock {\em arXiv.org}, November 2011.

\bibitem{Joulain:2003hc}
Karl Joulain, R{\'e}mi Carminati, Jean-Philippe Mulet, and Jean-Jacques
  Greffet.
\newblock {Definition and measurement of the local density of electromagnetic
  states close to an interface}.
\newblock {\em Physical Review B}, 68(24), December 2003.

\bibitem{Tersoff:1985wm}
J.~Tersoff and D.R. Hamann.
\newblock {Theory of the scanning tunneling microscope}.
\newblock {\em Physical Review B}, 31(2):805--813, January 1985.

\bibitem{Mulet:2001kp}
Jean-Philippe Mulet, Karl Joulain, R{\'e}mi Carminati, and Jean-Jacques
  Greffet.
\newblock {Nanoscale radiative heat transfer between a small particle and a
  plane surface}.
\newblock {\em Applied Physics Letters}, 78:2931--2933, May 2001.

\bibitem{Knoll:2000wm}
Bernhardt Knoll and Fritz Keilmann.
\newblock {Enhanced dielectric contrast in scattering-type near-field optical
  microscopy}.
\newblock {\em Optics Communications}, 182:321--328, August 2000.

\bibitem{Sun:2007cl}
Jin Sun, P~Scott Carney, and John~C Schotland.
\newblock {Strong tip effects in near-field scanning optical tomography}.
\newblock {\em Journal of Applied Physics}, 102(10):103103, 2007.

\bibitem{Joulain:2010bq}
Karl Joulain and J{\'e}r{\'e}mie Drevillon.
\newblock {Noncontact heat transfer between two metamaterials}.
\newblock {\em Physical Review B}, 81(16), April 2010.

\bibitem{Lax:1951tb}
Melvin Lax.
\newblock {Multiple Scattering of Waves}.
\newblock {\em Reviews of Modern Physics}, 23:287--310, October 1951.

\bibitem{Sipe:1987td}
J.E. Sipe.
\newblock {New Green-functions formalism for surface optics}.
\newblock {\em J. Opt. Soc. Am B}, 4:481--489, April 1987.

\bibitem{Chapuis:2008kca}
Pierre-Olivier Chapuis, Sebastian Volz, Carsten Henkel, Karl Joulain, and
  Jean-Jacques Greffet.
\newblock {Effects of spatial dispersion in near-field radiative heat transfer
  between two parallel metallic surfaces}.
\newblock {\em Physical Review B}, 77(3), January 2008.

\bibitem{Chapuis:2008kcb}
Pierre-Olivier Chapuis, Marine Laroche, Sebastian Volz, and Jean-Jacques
  Greffet.
\newblock {Radiative heat transfer between metallic nanoparticles}.
\newblock {\em Applied Physics Letters}, 92(20):201906, 2008.

\bibitem{Raether:1988ty}
Heinz Raether.
\newblock {\em {Surface Plasmons on Smooth and Rough Surfaces and on
  Gratings}}, volume 111 of {\em Springer Tracts in Modern Physics}.
\newblock Springer-Verlag, Berlin, 1988.

\bibitem{Joulain:2008tp}
Karl Joulain.
\newblock {Radiative Transfer on Short Length Scales}.
\newblock {\em Topics in Applied Physics}, 107:107--131, March 2008.

\bibitem{BenAbdallah:2011tz}
P~Ben~Abdallah, F.S.S. Rosa, M.~Tschikin, and S~A Biehs.
\newblock {Radiative cooling of nanoparticles close to a surface}.
\newblock {\em arXiv.org}, December 2011.

\bibitem{Huth:2010fg}
O~Huth, F~R{\"u}ting, S~A Biehs, and M~Holthaus.
\newblock {Shape-dependence of near-field heat transfer between a spheroidal
  nanoparticle and a flat surface}.
\newblock {\em The European Physical Journal Applied Physics}, 50(1):10603,
  March 2010.

\end{thebibliography}

\end{document}